# Interactions between syn-rift magmatism and tectonic extension at intermediate rifted margins

Peng Yang[a,b]*, Marta Pérez-Gussinyé[a]*, Shaowen Liu[b]*, Javier García-Pintado[a] and Gudipati RaghuRam[a]

(a) Marum - Center for Marine Environmental Sciences, Bremen Universität, Bremen, Germany
(b) School of Geography and Ocean Science, Collaborative Innovation Center of South China Sea Studies, Nanjing University, Nanjing, China
(*) Corresponding author. Email: pyang@uni-bremen.de; mpgussinye@marum.de; shaowliu@nju.edu.cn

**Abstract**

Intermediate rifted margins exhibit neither seaward dipping reflectors nor exhumed mantle at the continent-ocean transition (COT). Instead, they transition into normal-thickness, magmatic Penrose-type oceanic crust, and thus diverge from the classic magma-rich and magma-poor end-member models. However, several intermediate margins, such as the South China Sea (SCS), display detachment faulting similar to magma-poor margins and magmatic underplating typical of magma-rich ones. How tectonics and magmatism interact in these intermediate environments is poorly understood. Here we use 2D numerical models to demonstrate that the elevated initial geotherm inherited from prior plate subduction in the SCS explains several key observations: an early phase of wide rifting, subsequent localization onto core complexes with substantial footwall magmatic intrusions, and eventual formation of normal igneous oceanic crust at break-up. Thermal weakening caused by syn-rift footwall magmatic intrusions facilitates lower crustal ductile flow, promoting the development of 'rolling-hinge' type detachment faults and exhumation of core complexes. These structures are associated with accelerated tectonic subsidence, which is later moderated by detachment-related doming, as observed in the SCS. Normal-thickness oceanic crust occurs after break-up, even under ultra-slow extension rates used in our simulations, highlighting the importance of inheritance in determining margin architecture, the spatio-temporal distribution of syn-rift magmatism, and the nature of the COT. This behavior contrasts sharply with magma-poor margins, where a cooler lithosphere and similar ultra-slow extension produce no syn-rift magmatism, leading instead to crustal embrittlement, mantle serpentinization and exhumation at the COT.

# 1. Introduction

During continental rifting, lithospheric thinning is progressively influenced by the interaction of tectonic, magmatic, surface and hydrothermal processes, ultimately leading to continental breakup and the onset of seafloor spreading (Peron-Pinvidic et al., 2019; Brune et al., 2023; Pérez-Gussinyé et al., 2023). As the lithosphere thins, the underlying asthenosphere is drawn upward and may partially melt due to decompression, producing magma volumes that depend on mantle potential temperature and extension rate (Bown and White, 1995). Dykes intruding into the lithosphere can reduce the need for large-scale tectonic forces to rupture thick, strong continental lithosphere (Buck, 2004). More importantly, recent magnetotelluric imaging reveals a close spatial correlation of fault systems and transcrustal magma ascent at active continental rifts (Dambly et al., 2023). Such fault-aligned, self-sustained magmatic systems show that rifting is driven by a complex interplay between mechanical stretching and magmatic upwelling. Analogue models, which simulate magmatic underplating as a low-viscosity body of constant temperature, suggest that magmatic underplating promotes strain localization, while rift kinematics, in turn, control the pattern of magma emplacement (Corti et al., 2003). However, the long-term thermal effects of magma intrusion on extensional deformation remain enigmatic.

Rifted margins are commonly categorized based on the degree of magmatic activity during extension. Magma-poor margins (e.g., West Iberia-Newfoundland margin, WIM) are characterized by rapid initial subsidence, large-scale faulting, and minimal volcanism, often exposing exhumed mantle with little to no syn-rift magmatism (Franke, 2013). In contrast, magma-rich margins (e.g., North Atlantic) show abundant seaward-dipping reflectors (SDRs), significant magmatic intrusions and underplating, and over-thickened oceanic crust (Franke, 2013). Between these end-members lie the intermediate cases, typically exemplified by the northern South China Sea (SCS) margin, where drilling confirmed a rapid transition from breakup to oceanic crust (Larsen et al., 2018). Although a broad spectrum of magmatic processes has been suggested to control rift architecture, thermal evolution and breakup style (Pérez-Gussinyé et al., 2023), the nature of feedbacks between magmatism and tectonic extension is incompletely understood.

A key structural feature widely documented across numerous rifted margins is low-angle normal (detachment) faults, which are regarded as a fundamental mechanism driving margin evolution. Detachments within the distal sections of hyperextended magma-poor margins typically present a 'listric' geometry (concave upward) and are associated with increasing coupling of the crust-mantle system (Pérez-Gussinyé and Reston, 2001). Instead, intermediate margins exhibit concave-downward detachments accompanied by core complexes (e.g., Deng et al., 2020; Xu et al., 2024), which start at depth with a relatively steep angle and progressively curve to a shallower dip towards the surface. Analogue and numerical models suggest that the latter structures require lithospheric mechanical decoupling, typically facilitated by a weak, easily flowing ductile lower crust (Brun et al., 1994), post-orogenic crustal thickening and high initial thermal gradients (Tirel et al., 2008), and/or a strong rheological contrast imposed by inherited lithological layering (Huet et al., 2011). Particularly, magma emplacement has been identified as a key driver of core complex formation and exhumation, through mechanisms such as stress rotation (Parsons and Thompson, 1993) and transient thermal perturbation (Lister and Baldwin, 1993). Field observations clearly show systematic magmatic and tectonic extension, with pluton emplacement and successive detachments development (e.g., Jolivet et al., 2021). Nevertheless, the genetic relationships among magmatism, detachment faulting, and core complex formation remain unclear due to the complexities of exhumation processes.

To address these issues, we focus on the northern continental margin of the SCS, the largest marginal sea in the western Pacific (**Fig. 1**). Following Late Mesozoic landward subduction of the Paleo-Pacific plate, Cenozoic rifting in the SCS proceeded in a wide-rift mode (Deng et al., 2020), ultimately accommodating over 600 km of extension across the conjugate margins. High-resolution seismic data reveal large-scale detachment faults that exhume middle/lower crustal rocks, forming core complexes from the proximal margin to the continent-ocean transition (COT) (e.g., Deng et al., 2020; Ding et al., 2020; Zhang et al., 2021a; Xu et al., 2024). In the distal domain, seismic and borehole data document high-velocity lower crust (HVLC), laccoliths, and sills, evidence of syn-rift magmatic underplating and intrusion (e.g., Zhang et al., 2023). While these features are typically associated with magma-rich margins, seaward-dipping reflectors (SDRs) are notably absent. The continental taper features titled fault-bounded blocks with wedge-shaped sediments, reminiscent of magma-poor margins (Zhang et al., 2024). However, unlike those margins, drilling indicates a rapid transition from continental rifting to seafloor spreading, without evidence of mantle serpentinization (Larsen et al., 2018; Zhang et al., 2024). These contrasting features suggest that the SCS differs prominently from magma-rich and -poor margins, likely due to subduction-related inheritance, a short period of thermal relaxation, and/or elevated late rifting, initial spreading rates and the absence of plume-related thermal anomaly (Xu et al., 2024; Zhang et al., 2024). While previous numerical studies have emphasized the roles of lithospheric rheology (Brune et al., 2017; Pérez-Gussinyé et al., 2020), far-field stress (Le Pourhiet et al., 2018), and pre-rift orogenic inheritance (Li et al., 2024) in shaping the SCS margin, none have explicitly incorporated melting processes.

Here, we present two-dimensional (2D) numerical models that couple mechanical deformation with surface processes, parametrized hydrothermal cooling, and magmatic processes to explore the evolution of the SCS margin. Specifically, we assess the impact of heat release by magmatic intrusion on extensional deformation, pressure-temperature-time (*P-T-t*) paths, and tectonic subsidence. Our results reproduce key features observed in the SCS—including a wide rift system characterized by late syn-rift detachments with large magmatic footwall intrusions and underplating, and the formation of normal oceanic crust at breakup—and reveal the feedbacks between magmatism and tectonics that shape the architecture of intermediate margins. We then compare these results with those from other magma-poor margins and propose a distinct deformation mode characteristic of magma-influenced, wide rift systems.

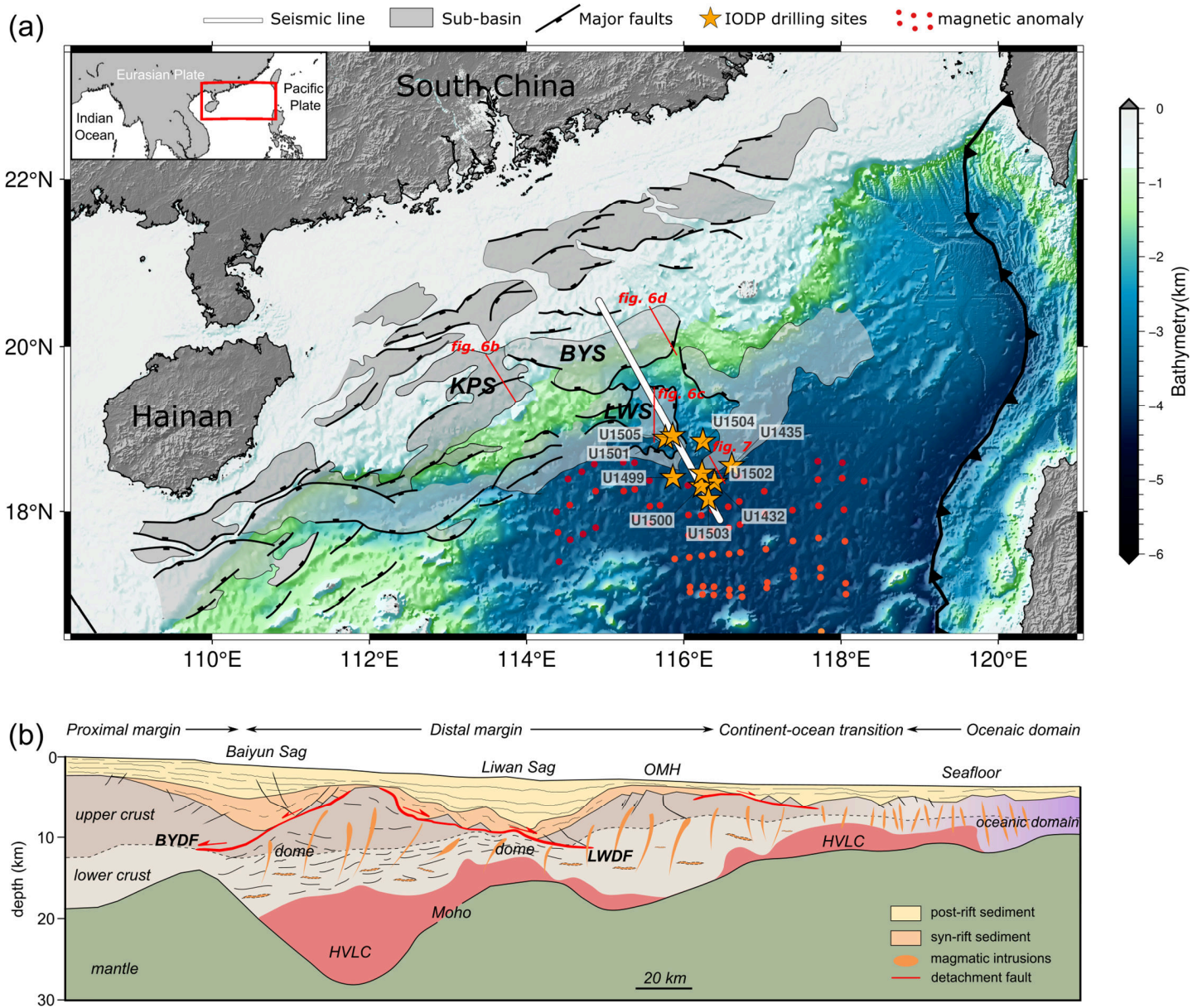

Fig. 1. (a) Bathymetric map of the northern continental margin of the South China Sea (SCS), showing major faults, sub-basins, and IODP drilling sites. Abbreviations: BYS – Baiyun Sag, LWS – Liwan Sag, KPS – Kaiping Sag. (b) Interpreted cross-section along the white line in (a), illustrating the main features of the rifted margin based on published seismic reflection/refraction profiles (Deng et al., 2020; Ding et al., 2020; Mohn et al., 2022; Zhang et al., 2023). Abbreviations: OMH – outer margin high, HVLC – high velocity lower crust, BYDF – Baiyun detachment fault, LWDF – Liwan detachment fault.

## 2. Method

We employ Rift2Ridge, a 2D finite-element code with non-Newtonian visco-elasto-plastic rheologies, to solve the momentum, mass, and energy conservation equations in incompressible Stokes flow (Supplementary Text S1, S2). The code is built upon MILAMIN mechanical and temperature solvers (Dabrowski et al., 2008) and has been further developed in previous studies (Pérez-Gussinyé et al., 2020; Raghuram et al., 2023; Mezri et al., 2024; García-Pintado and Pérez-Gussinyé, 2025). This enhanced model incorporates strain softening, surface erosion and sedimentation, parametrized hydrothermal cooling, mantle serpentinization, and melting processes (Supplementary Text **S3–S6**). The model domain is 400 km wide and 150 km deep. We apply a Winkler boundary condition at the model base to ensure isostatic equilibrium and a stress-free surface to simulate dynamic topography. Extension is initiated at the model center using a randomly distributed weak seed (Supplementary Text **S3**). Detailed model setup and justification of initial thermal and rheological conditions are described in Supplementary Text **S7,** Table **S1,** and illustrated in **Fig. 2**.

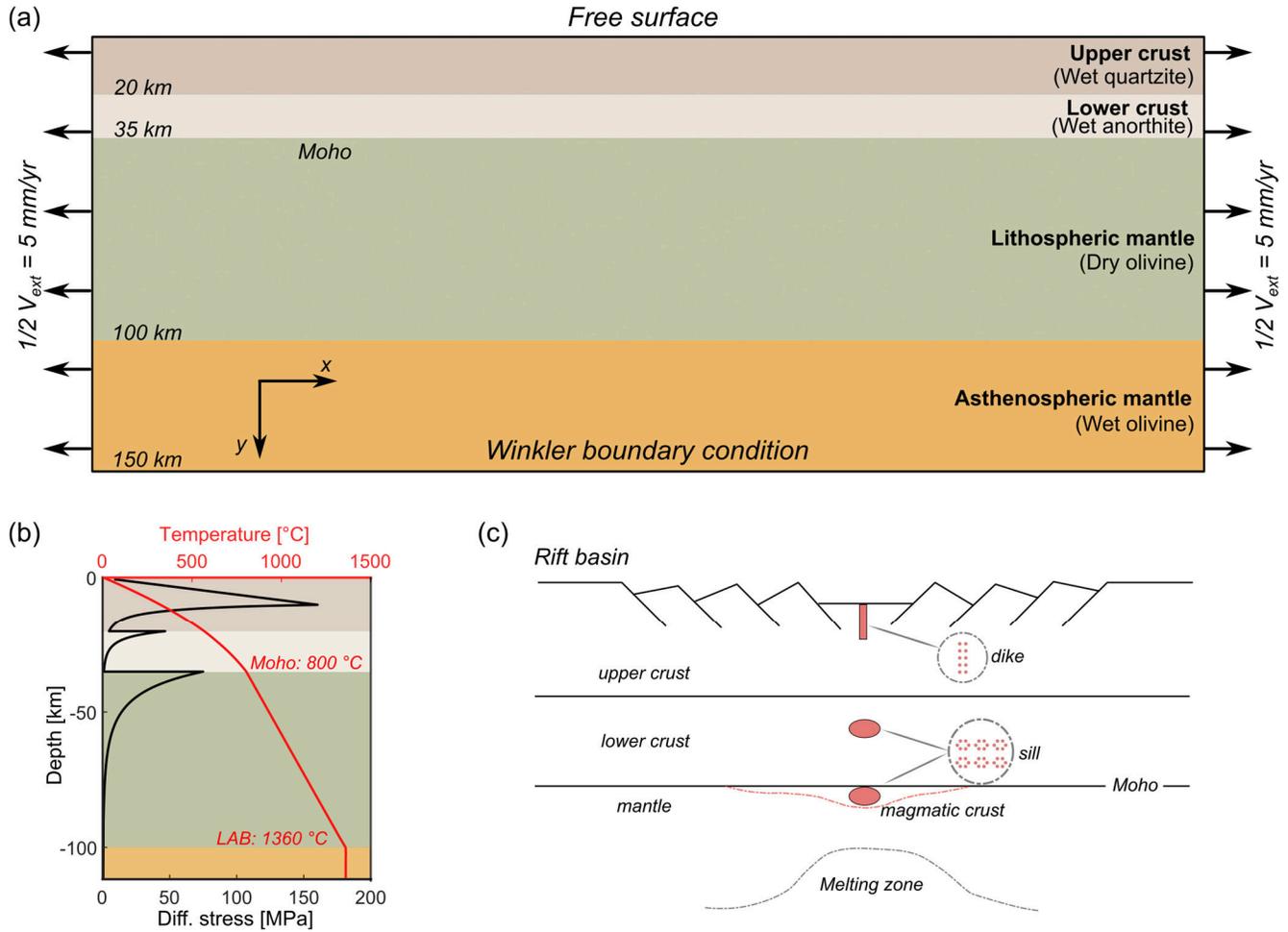

Fig. 2. (a) Illustration of the model setup, including the geometry, layer properties, and boundary conditions for the reference model. (b) Initial temperature and strength profiles, assuming a strain rate of $1 \times 10^{-15}$ s$^{-1}$. (c) Conceptual model for simulating melt intrusion and underplating. Dykes and sills are represented as track points carrying magma. Drawing is not to scale.

## 2.1 Melting processes

As extension proceeds, mantle upwelling drives decompression melting once the geotherm intersects the dry solidus. The melting process consumes latent heat via fusion and locally reduces temperature. Mantle partial melting is modeled following Phipps Morgan (2001) (Supplementary **Text S5.1**), assuming a reversible adiabatic process. Mantle depletion is updated at each time step based on melt fraction, and the new depletion level is used to adjust the solidus temperature accordingly. As mantle depletion increases from 0% to 4%, the rheology of partially melted mantle transitions linearly from wet to dry olivine and is modeled as a compositional mixture of both. Mantle depletion increases viscosity and lowers density relative to the undepleted mantle matrix (Supplementary **Text S5.1**). Emplacement of melts releases latent heat (via crystallization) and sensible heat (via cooling) to surrounding rocks upon emplacement (Supplementary **Text S5.2**).

The mechanisms of melt migration and their coupling with deformation are complex, but it is widely agreed that melt is generated over a broad region within the mantle and then travels laterally uphill towards the ridge center, from where it ascends to form new crust (e.g., Sparks and Parmentier, 1991). In extending rift systems, the top of this melting region and the base of the crust are usually farther apart and likely cooler than at mid-ocean ridges, creating pressure and temperature conditions that may favor stress-driven melt segregation. This is supported by numerical simulations (e.g., Katz et al., 2006), experimental rock deformation studies (Kohlstedt and Holtzman, 2009), and geological investigations (e.g., Kelemen and Dick, 1995; Kruckenberg et al., 2013), all indicating that melt predominantly rises through mantle shear zones linking the melting region to the crust. Magnetotelluric imaging beneath the Main Ethiopian Rift also reveals a clear deflected magma conduit from the melting zone into the crust along shear zones (Dambly et al., 2023). Accordingly, our simulations parametrize melt migration by assuming that mantle melts are first emplaced beneath the Moho, where major mantle shear zones intersect and strain rates are highest.

At the crustal level, the depth of melt emplacement is governed by structural heterogeneities, the brittle-ductile transition, and local stress conditions (Parsons et al., 1992). Geochemical evidence shows significant focused mid-crustal magma intrusions in addition to those straddling the Moho (Wong et al., 2023). An emerging conceptual model describes a transcrustal, interconnected magmatic system characterized by magma conduits and unstable mush reservoirs (Cashman et al., 2017). However, modeling such magmatic systems remains challenging due to the complex multiphase chemistry and physics involved, particularly regarding melt trajectories and emplacement. For simplicity, we model melt intrusion through an idealized, kinematic framework, with magma emplaced as instantaneous pulses (**Fig. 2c**), due to the relatively rapid timescale of dyke and sill emplacement compared to lithospheric thermal evolution (e.g., Rubin, 1995). Accumulated melts at the crustal base form a ponding zone, analogous to the HVLC, with subsequent batches emplaced beneath this new magmatic base. Remaining melts ascend into the lower crust, where emplacement follows a Gaussian distribution, and intrude beneath the top basement at each timestep. Melts are modeled as sills in the uppermost mantle and lower crust, and as dykes in the upper crust (**Fig. 2c**). Recent numerical modeling has revealed the major control of crustal stresses on magma pathways, with dyke propagation shown to occur perpendicular to the minimum compressive stress (Maccaferri et al., 2014; Ferrante et al., 2024). Specifically, the location of rift-related magmatism can shift between in-rift and off-rift zones, determined by the evolving graben topography as the rift matures (Ferrante et al., 2024). However, most magmatic additions in the SCS margin appeared during the late rifting stage (Zhang et al., 2021b), unlike other margin settings affected by mantle plume (e.g., Pelotas, Santos, and Campos basins), where dyke swarms emerge at the rifting onset (Jackson et al., 2000). The absence of initial melts in our models is therefore consistent with the SCS margin, and subsequent sedimentation could diminish topographic variation, potentially bypassing or shortening the early off-rift magmatism stage. Additionally, because melt generation mainly occurs when the crust thins below 20 km (see Results), lateral deflection of ascending melts from deeper ponding zones within the lower crust is limited. Therefore, although our models do not capture horizontal magma shifting, this limitation does not compromise the validity of our first-order conclusions.

Estimating the partitioning of melt within the crust is also challenging in natural volcanic systems. Compilations from continental volcanic regions report a wide range of intrusive-to-extrusive ratios, ranging from 3:1 to 10:1 (White et al., 2006), although these results do not exhibit systematic variations across different volcanic settings. Such estimates remain uncertain because of the potential erosion of extruded volcanic rocks and the limitations of seismic imaging in capturing long-term melt volume. Here we assume 1/3 of the total produced melt intrudes beneath the top basement, while the rest is split into 1/3 and 2/3 emplaced in the lower crust and beneath the Moho, respectively. We assess the sensitivity of this assumption in a test model (Model 7, Supplementary **Table S2**).

*2.2 Model limitations*

This study represents a first-order approximation of magma-tectonic interactions affected by magmatic heat release. Key simplifications include: (1) The 2D numerical domain does not capture 3D along-strike margin variability (Ding et al., 2020). Also, the thermal impact of magmatic bodies may be overestimated in 2D, where intrusions are modeled as infinite sheets rather than finite 3D volume. (2) The model is incompressible, meaning that we do not account for crustal thickening by magmatic additions (i.e., no mesh accommodation for melts). This leads to an underestimation of final crustal thickness, as magmatic additions contribute ~20% of the total crustal thickness in the northern SCS margin (Zhang et al., 2021b). (3) We omit the complexities of a multiphase melting system and exact magma migration trajectories controlled by the stress field. Future work on continental rifting simulations will require more physics-based melt migration models (e.g., Pusok et al., 2025).

## 3. Results

We first present four numerical models (Models 1–4) to investigate how melt generation and emplacement influence rift dynamics. All models employ a half-extension velocity of 5 mm/yr, an initial Moho temperature of 800°C, and a crustal thickness of 35 km. Building on Model 4, we further explore the effects of lithospheric strength by modifying initial Moho temperature (Models 5a–5b) and the ratio of upper/lower crust thickness (Models 6a–6b). Initial Moho temperature is adjusted by altering lower crustal heat production. Model 7 explores the sensitivity of melt volume partitioning. Full model configurations are described in the Method section and Supplementary **Table S2**.

## 3.1 Melt-absent and melt-present models

Model 1 (melt-absent) serves as the base model (**Fig. 3a,** Supplementary **Movie S1**). During early stretching phase, high-angle normal faulting distributes in the brittle upper crust, forming widespread horst-graben structures. Due to lithostatic pressure gradients, the lower crust undergoes slight uplift beneath areas of pronounced upper crust deformation. By 16 Myr, strain localizes into several sets of high-angle faults and shear zones. As slip accumulates, early faults rotate to lower angles, and new high-angle faults form on the hanging wall of the breakaway surface, which resembles the rolling-hinge model (Buck, 1988). Core complexes emerge as deep rocks are exhumed, accompanied by small overriding blocks situated above the footwall. Lithospheric thinning is depth-dependent due to decoupling between the viscous lower crust and mantle, leading to earlier lithospheric mantle breakup (~21 Myr) relative to the crust. Between 22–26 Myr, successive phases of polyphase faulting develop, during which young faults transect earlier detachment footwall, creating zones of newly accreted basement with undulating, low-angle fault surface (-20 to 10 km in the model section). This phase is accompanied by the development of a new concave downward fault (-28 to -15 km at 23 Myr) to the left of the polyphase faulting region. By 28 Myr, another large concave downward fault forms (25–40 km) but is disrupted by later faulting. When the crust thins to ~10 km, conductive cooling induces crustal embrittlement, promoting rapid strain localization through faulting. Crustal breakup and seafloor spreading occur at 33 Myr. The final structure is an asymmetric wide margin with a relatively thinner crustal section within its distal domain on the left.

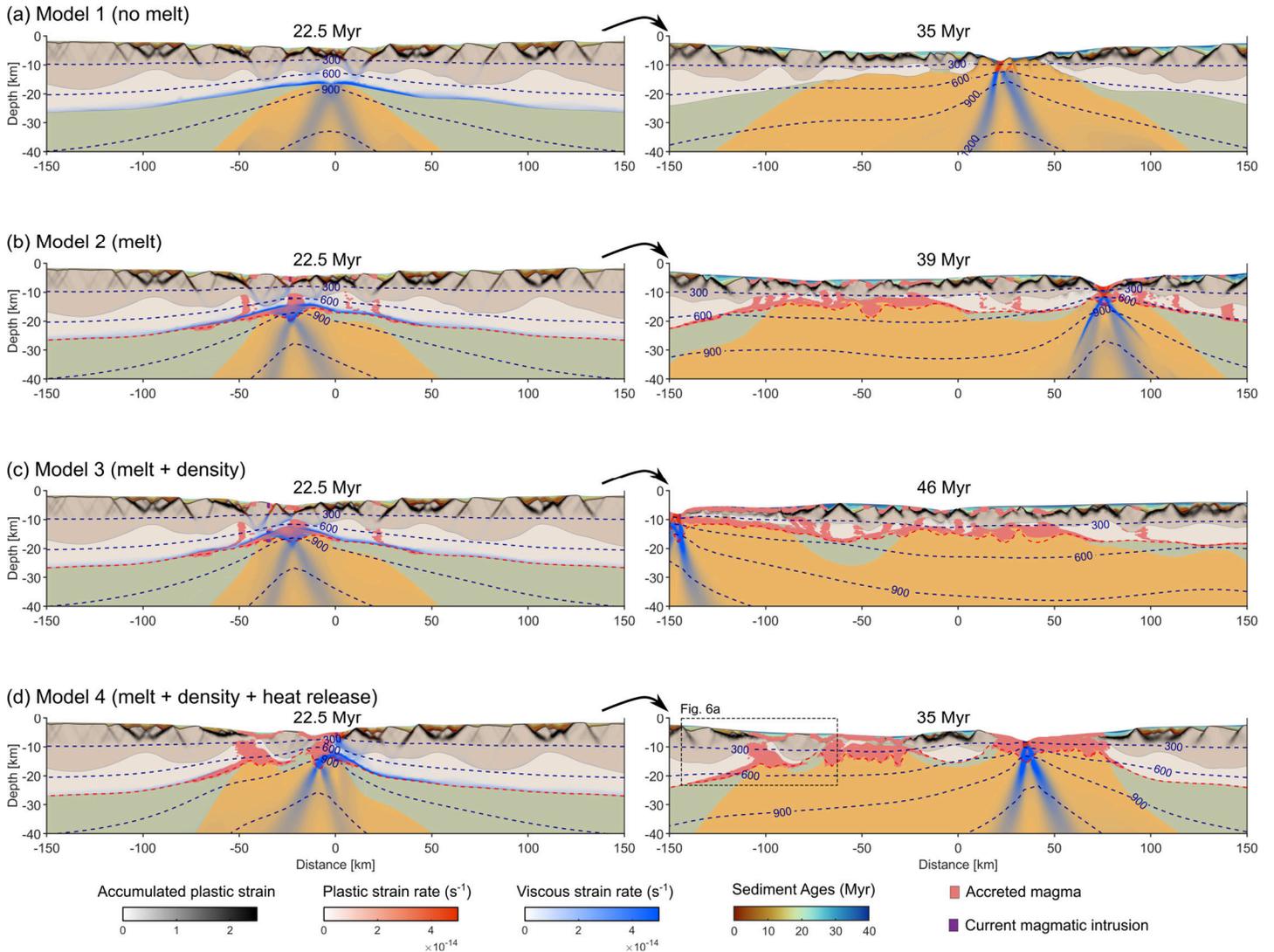

Fig. 3. Snapshots of the rift architecture at 22.5 Myr (left panel) and 35 Myr (right panel) showing the effects of melt emplacement. All models assume an extension velocity of 10 mm/yr, an initial Moho temperature of 800°C, and a mantle temperature of 1360°C. Model 1 is the base model, excluding any magma-related processes. Model 2 incorporates mantle partial melting but omits the effects of melt density and heat release. Model 3 accounts for the density effect of melt, while Model 4 fully integrates all magmatic processes, which serves as the reference model. Blue dashed lines are isotherms, arranged from top to bottom as 300, 600, 900, and 1200°C. Red dashed lines mark the base of magmatic crust.

Model 2 includes mantle melting via decompression but excludes thermal and density effects of melt emplacement (**Fig. 3b,** Supplementary **Movie S2**). The only difference of this model from Model 1 lies in that the depleted mantle acquires an increased viscosity and decreased density due to melting (Supplementary **Text S5.1**). This model follows the same stretching phase as Model 1 until asthenospheric upwelling induces decompression melting at 12.6 Myr. Transition from wet to dry olivine increases the viscosity of the melt zone. Mantle strengthening progressively inhibits further strain localization in the initially active melting region, shifting deformation away from the original rift center to areas with non-depleted mantle. This leads to strain migration in the crust, and successive diachronous faulting segments an early detachment fault (-75 to -50 km at 27 Myr) into half-graben systems. By 35 Myr, deformation continues to migrate, culminating in crustal breakup at 39 Myr. Compared to Model 1, the distributed crustal faulting results in a prolonged rifting phase, a wider margin, and a thinner crust. Faults in this case lack explicit migration direction and exhibit an out-of-sequence development.

Model 3 (**Fig. 3c**, Supplementary **Movie S3**) incorporates melt density variations by mixing melts with the original background material according to melt fraction. The margin architecture in this model exhibits a breakup position distinct from that of Model 2, which resulted from long-distance rift migration during the late extension stage. Apart from this, the faulting and magmatism patterns remain generally comparable, implying a minimal effect of melt density on rifting dynamics (**Movies S2, S3**). In both models, the crust fails to break up by 30 Myr, which is inconsistent with the SCS rifting duration, making these models unsuitable.

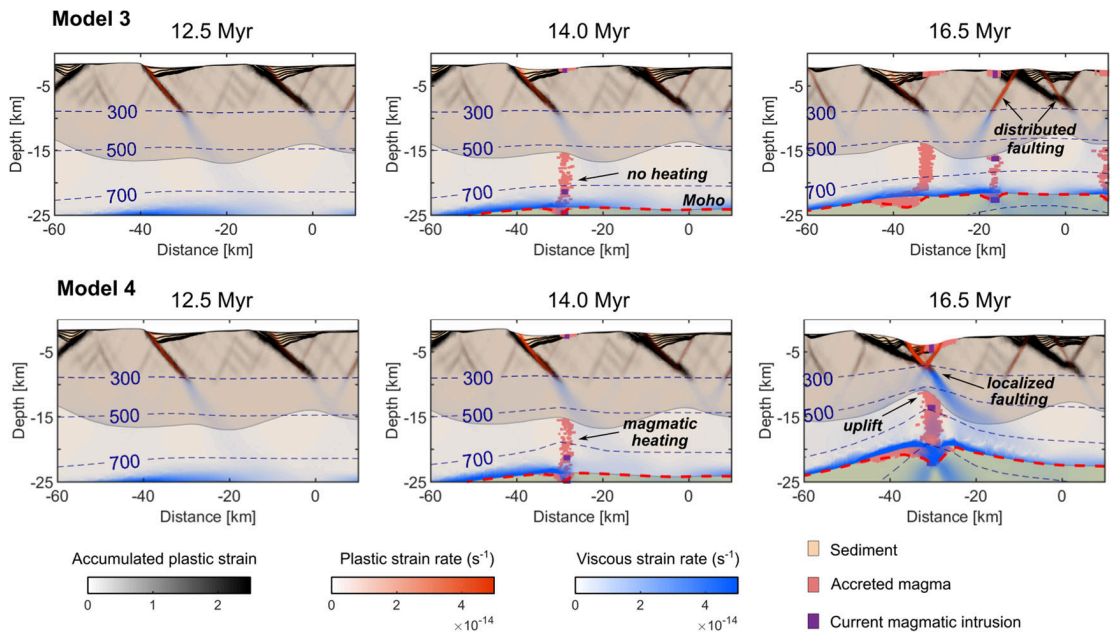

Fig. 4. Comparison of deformation evolution between Model 3 (excluding magmatic heat release) and Model 4 (incorporating heat release). Blue dashed lines are temperature isotherms.

Model 4, the focus of this study, incorporates melt generation, density effects, and crucially, heat release from melt emplacement (**Fig. 3d, Movie S4**). Melt production begin at 12.6 Myr, as in Models 1–3. Brittle faults in the uppermost crust propagate as a ductile shear zone along the upper-lower crust interface. Unlike Model 3, magmatic heat localizes strain in the upper crust (**Fig. 4**). As extension and melt addition continue, the heated footwall uplifts via isostatic rebound and each new fault in the hanging wall rotates to a low angle before being abandoned as a new fault develops. This produces a concave-downward detachment fault, with its largest segment expressed as ductile shear band rooting in the deep crust. By 19.4 Myr, footwall cooling and mantle strengthening surpass magmatic weakening, triggering rift migration (**Movie S4**). By 22.4 Myr, a second detachment forms between -20 km and 10 km, which is later segmented by another counter-dipping detachment fault. Melt-facilitated strain re-localization leads to final crustal breakup at 31.2 Myr. A detailed analysis of the detachment cycle is given in section 4.2. Compared with Models 1–3, strain localization aided by magmatic heating in Model 4 extends the length (~40 km) and lifespan of the first detachment, allowing intense lower-crustal exhumation at shallower levels. Some detachment faults remain intact without being disrupted by later polyphase faulting, and rift migration proceeds consistently oceanward, unlike the out-of-sequence behavior in Models 2–3. Note that although the final margin architecture shows asymmetric melt distribution on both conjugate margins, this reflects rift migration toward thicker and weaker crust (i.e., toward one side of the original model center) rather than large-scale asymmetric melt emplacement (**Movie S4**).

*3.2 Additional models*

Models 5a and 5b vary initial Moho temperatures (600°C and 740°C, respectively) (**Figs. 5b–5c**), while considering melting and heat release from melt emplacement as in Model 4. In Model 5a, the colder lithosphere develops large-offset normal faults, with faulting rapidly migrating basinward without exhumation, leading to quick breakup and narrow symmetric conjugates. In contrast, relatively hotter lithosphere (Model 5b) extends in wide rift mode with detachment faulting and lower crustal exhumation similar to Model 4. However, stronger crust-mantle coupling produces narrower conjugate margins and earlier breakup. Magmatism migrates laterally only once after the first detachment formation, and the long-lived polyphase faulting seen in Models 2–3 is absent.

Models 6a and 6b examine different upper/lower crustal thickness ratios, keeping the total crustal thickness at 35 km and all other parameters identical to Model 4. In Model 6a (15 km upper crust), reduced radioactive heating and stronger coupling localize plastic deformation near the upper-lower crust interface (**Fig. 5d**). Extension is accommodated primarily by closely spaced upper crust faults, with limited lower crust involvement. With continued extension, the upper crust undergoes intensive thinning, exhuming lower crust, followed by rapid strain localization and breakup at 19.1 Myr. Conversely, in Model 6b (25 km upper crust; **Fig. 5e**), reduced crustal strength and increased geothermal gradient result in prolonged upper crustal stretching. The delay in strain re-localization ultimately postpones crustal breakup with respect to Model 6a, occurring at 30.7 Myr. Despite a thin lower crust, this model still predicts detachment and core complex formation facilitated by melt intrusion.

Model 7 increases melt emplacement at depth by reducing upper crustal intrusion from 1/3 to 1/4, producing a thicker underplated body (**Fig. 5f,** between the dotted red line and the Moho). Rift evolution resembles Model 4, with detachment fault formation, lower crust exhumation, and oceanward rift migration. However, greater thermal weakening in deep levels accelerates strain localization during the second-phase rift migration, leading to an earlier breakup (27 Myr).

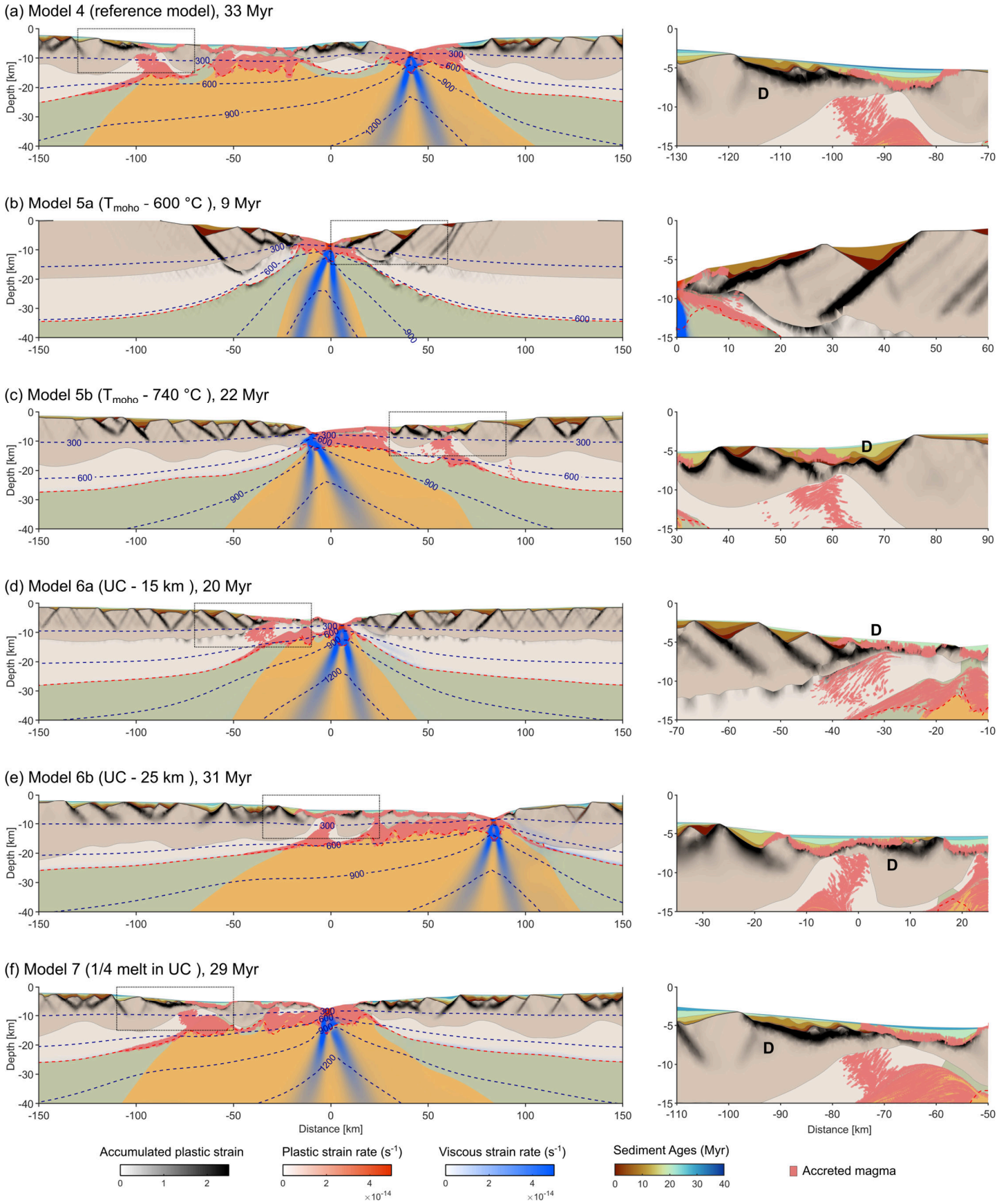

Fig. 5. Parametric analysis based on (a) Model 4, examining the effects of initial Moho temperature, crustal thickness distribution, and melt volume partitioning. (b-c) Model 5a and Model 5b test initial Moho temperatures of 600 °C and 740 °C, respectively. (d-e) Model 6a and Model 6b examine variations in the upper-to-lower crust thickness ratio while maintaining a constant total crustal thickness of 35 km, with an initial upper crust thickness of 15 km in Model 6a and 25 km in Model 6b. UC – upper crust. (f) Model 7 explores melt volume distribution, with 1/4 of the total generated melt intruding into the upper crust. The right-panel figures are enlarged views of the boxed area in the left panel, showing the detachment structure. D-detachment.

Overall, despite differences in parameterization, all models that incorporate high initial Moho temperature develop detachment faults and core complex systems with significant lower crustal exhumation. Early studies suggested that core complex formation requires a pre-thickened orogenic crust (> 45 km) and a high Moho temperature (> 800 °C) (Tirel et al., 2008). This view has been challenged by the influence of inherited lithological layering or the presence of a rheologically weak unit, which permits a lower Moho temperature required for lower crustal exhumation (Huet et al., 2011). In our models, a similarly low-viscosity zone can be produced by magmatic heating, promoting the development of detachment systems in non-over-thickened continental crust. Importantly, detachment faults in magmatic models (**Fig. 5**) persist throughout the rifting history without being overprinted by polyphase faulting, enabling sustained lower crustal exhumation that reaches the surface. These results underscore the essential role of magmatism in promoting and preserving large-scale detachment fault systems.

## 4. Discussion

### *4.1 Comparison between model predictions and seismic profiles*

Model 4 successfully replicates key features observed in the northern SCS continental margin, including wide rifting, seaward migration of magmatic emplacement (Zhang et al., 2021b), hyper-extended crust, and boudinage structures (Deng et al., 2020) (**Fig. 1b**). In particular, seismic data from the Liwan, Kaiping and Baiyun sags reveal notable detachments and core complex structures marked by undulating internal reflectivity, which are comparable to the modeled lower crustal deformation fabrics (**Fig. 6**). These structures are interpreted as local anomalies within wide rift mode (Deng et al., 2020), which can be attributed to rheological heterogeneities within the ductile lower crust caused by magma intrusion in our models. A similar crustal architecture is observed in the Campos Basin, where exhumed lower crust is overlain by SDRs (Alvarez et al., 2024). Our models also reproduce both continent-ward (**Fig. 5e**) and seaward-dipping (**Fig. 5a**) detachment faults, with their dip direction determined by the orientation of the precursor normal faults (Supplementary **Movie S4**). Magmatic bodies are predicted to accrete in detachment footwalls, forming large underplated pockets (**Fig. 6a**), consistent with high-velocity layers observed in seismic refraction profiles (**Fig. 1b**) (Zhang et al., 2023). Such continuous magmatic ponding indicates a clear spatial link between magmatism and detachment systems. In contrast, the initial stage of continental core complex formation is observed in the proximal domain of the SE SCS, where high-amplitude basement reflectors remain isolated from syn-rift sediments (Legeay et al., 2024). This structure resembles the core complex in our magma-absent model (Model 1, **Fig. 3a**), where lower crustal exhumation does not reach the surface, leading to an immature stage development.

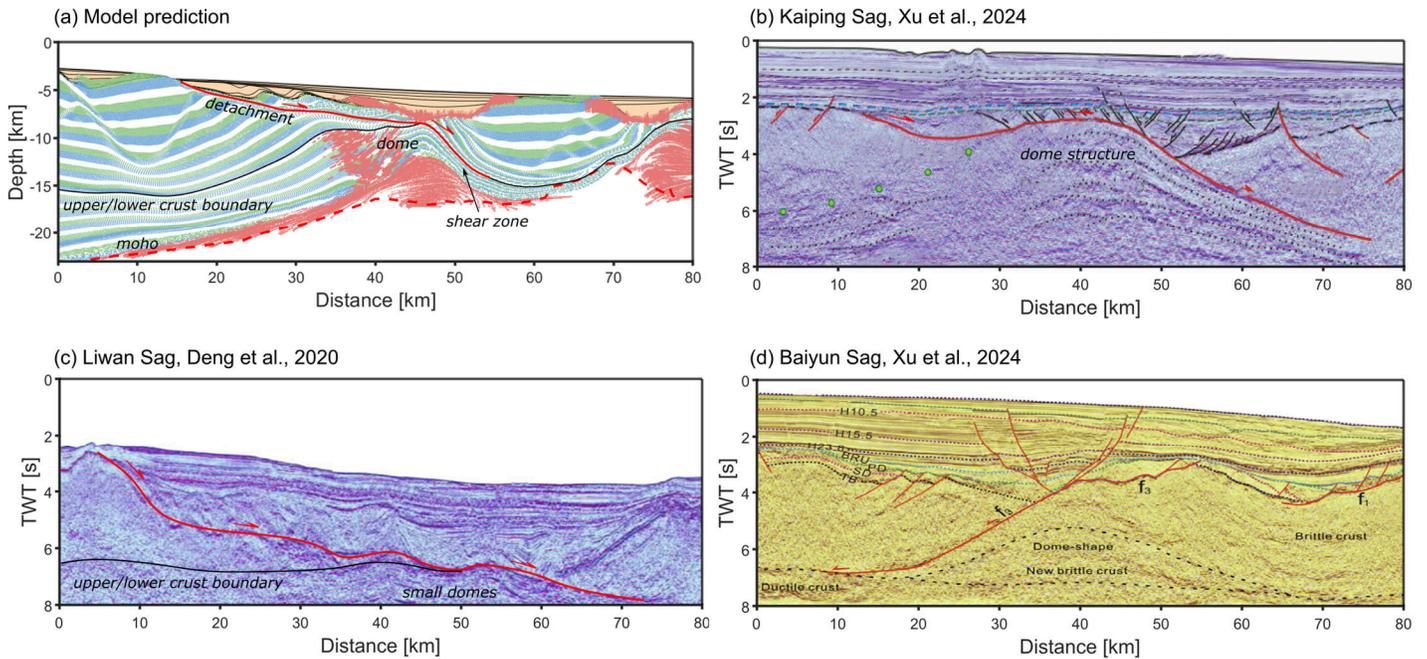

Fig. 6. Comparison of modeled and observed structural features in sub-basins of the northern SCS margin. The model and seismic profiles are presented at the same scale. (a) Results of Model 4 at 42 Myr, with colored tracers set in the beginning to show the crustal deformation. Pink points are melt tracers. (b-d) Structure and interpretations from published seismic profiles from Kaiping, Liwan and Baiyun sags (see **Fig. 1a** for basin and seismic profile locations) (Deng et al., 2020; Xu et al., 2024). Crustal reflectors reveal remarkable upward doming core-complex geometry and low-angle detachment faults intersected by steeper faults or shear bands penetrating the lower crust. The modelled detachment best resembles that observed in the Kaiping Sag (b). Still, model and data can not be compared one to one, as detachments in nature are 3D and the geometry is presented in two-way travel times (TWT) in seismic data.

Detachment fault and core complex structures occur not only within the sags of hyper-extended distal domain but also in the COT zone (Zhang et al., 2021a; Mohn et al., 2022), where they affect final lithosphere separation. Drilling data reveal the emplacement of mid-ocean ridge basalts and thermal alteration of syn-rift gravels, demonstrating the involvement of syn-rift magmatism in the breakup process (Larsen et al., 2018; Nirrengarten et al., 2020). Such tectono-magmatic architecture of the COT, as imaged in seismic profiles, is well reproduced by our model (**Fig. 7**), where increasing melt supply transitions the system from detachment-controlled to normal igneous crustal accretion (approximately 5–6 km thick).

Numerical modeling by Li et al. (2024) suggests that exhumation at the SCS margin could be facilitated by pre-existing thrust faults or inherited structures, but detachment faults in their models are considerably shorter than those observed. This discrepancy implies that additional factors, potentially including prolonged strain localization related to heating and weakening due to magmatic emplacement, as captured in our reference model, may play a crucial role in the development of large-scale detachment faults in natural settings.

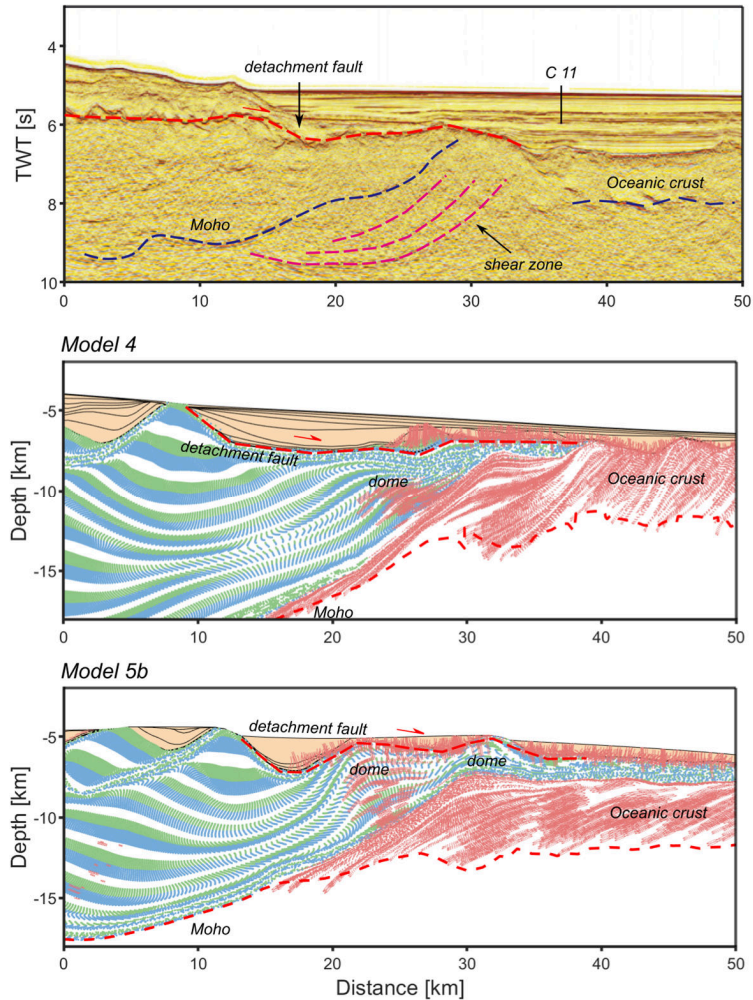

Fig. 7. Comparison of the observed seismic profile (Zhang et al., 2021a) from the continent-ocean transition in the northern SCS margin and model predictions (Model 4 and Model 5b). Markers shown in model predictions are the same as those of **Fig. 6**. See Fig. 1 for the location of the seismic profile. The breakup structure is interpreted as a core complex-like detachment system, which results in the exhumation of footwall rocks and accreted magma. Note that the horizontal orientation in this figure is reversed compared to **Fig. 5**.

*4.2 The interaction between magmatism and deformation*

Previous analogue models, which pre-assign magma underplating as a low-viscosity body (e.g., Corti et al., 2003), demonstrate strong coupling between deformation and magma emplacement, with viscous magma influencing local fault patterns and promoting core complex development. However, these models neglect magmatic heating and associated thermal weakening of the crust. Our numerical models generate decompression melts within a self-consistent rifting framework and incorporate the thermal effects of magma emplacement, which provide new insights into how deformation and magmatism interact.

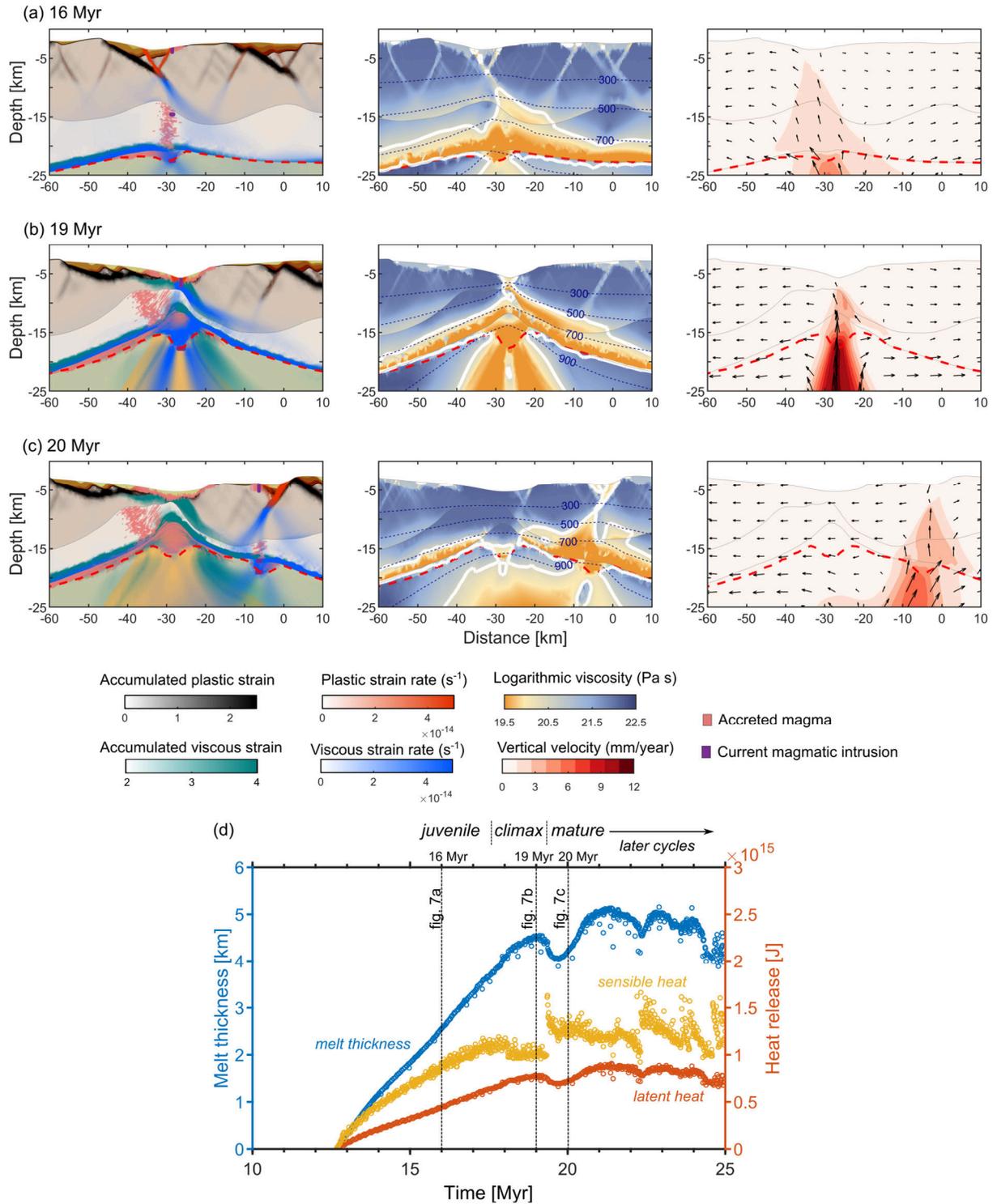

Fig. 8. Evolutionary stages of the reference model (Model 4). Left panels: accumulated plastic strain and strain rate. Middle panels: viscosity field with white iso-viscosity contours ($10^{20.5}$ Pa s) and blue isotherms. Right panels: velocity vectors (black arrows) superimposed on vertical velocity magnitude (colored background). Red dashed lines denote the base of the magmatic crust. (a) *Juvenile* stage at 16 Myr. Localized plastic strain accumulation above the melt emplacement zone. A low-viscosity crustal pocket develops at the upper crustal fault tip, connecting to a deep shear band, accompanied by lower crustal doming. (b) *Climax* stage at 19 Myr. High-angle normal fault rotation into low-angle detachment fault facilitates exhumation of lower crustal materials to the surface. Vertical mantle flow attains its maximum value. (c) *Mature* stage at 20 Myr. Crustal cooling generates horizontal strength contrasts, driving rift migration. (d) Temporal evolution of generated melt thickness, sensible and latent heat released from magma emplacement. Note that here melt

thickness represents the total melt volume extracted from the mantle, while the pink magmatic layers (between the red dotted line and the Moho) plotted in other figures (e.g., **Fig. 3**) correspond to the melts emplaced beneath the Moho. Latent heat release exhibits oscillation correlated with melt thickness, indicating its direct dependence on melt volume. In contrast, sensible heat release displays discontinuous jumps resulting from pronounced temperature differences between the emplaced melt and the host rock (**Supplementary Text S5.2**). Toward the end of the detachment, when the faulting zone has been heated by previously emplaced melts, elevated ambient temperatures inhibit sensible heat release. Following rift migration, melts intrude into cooler, previously unheated regions, resulting in greater sensible heat release. These abrupt shifts signal the activation of a new fault and the initiation of new detachment-melt cycles.

In this section, we interpret the detachment fault cycle using a three-phase framework (Handy et al., 2001) and force-balance model (Lavier et al., 2000; see also review in Pérez-Gussinyé and Liu, 2022). This model assumes that continued fault slip depends on two components: bending stress $F_B$, and the force to sustain the fault slip $F_S^{fault}$. A fault can remain active as long as $F_B < F_S^{new} - F_S^{fault}$, where $F_S^{new}$ is the force required to initiate a new fault. Here, $F_B$ increases with fault offset and the square of thickness of the brittle layer, while $F_S^{new}$ depends on the initial cohesion and frictional angle, and $F_S^{fault}$ is controlled by the degree of strain weakening along the existing fault plane.

During the *juvenile* phase (**Fig. 8a**), incipient melt emplacement raises isotherms and generates a convex-upward thermal structure. Brittle strain localizes along a pre-existing normal fault and continues as a viscous shear band that extends into the lower crust. This is consistent with experimental studies showing that rock strength decreases exponentially with increasing melt volume, facilitating strain localization in areas of melt emplacement (Handy et al., 2001). At this stage, the thinning brittle layer reduces $F_B$ and, increasing strain softening lowers $F_S^{fault}$. As a result, the weakening processes dominate the total force balance, allowing the fault to accumulate large offset. In the *climax* phase (**Fig. 8b**), viscosity reduction facilitates lower crustal flow beneath the low-angle fault, accompanied by the development of core complex. This suggests that magmatic heating alone can promote core complex formation through thermal weakening, even without explicitly modeling viscous mechanics of magma (Corti et al., 2003). Vertical mantle flow velocity increases, enhancing melt production and heat release (**Fig. 8b**). Increasing melt emplacement further promotes strain localization, maximizing footwall exhumation and detachment flattening. Peak vertical mantle flow velocity and maximum melt production coincide with this stage (**Fig. 8d**). In the final *mature* phase, the extension on the detachment is considerable, and the mantle preferentially uplifts towards the tip of the detachment at depth (**Fig. 8c**), so that the footwall tends to cool and the hanging wall warms up. Once the detachment has accumulated a large offset, the increase in $F_B$ makes continued slip less favorable, leading to rift migration and onset of a new detachment cycle (Supplementary **Fig. S1**). Meanwhile, vertical mantle flow velocity and melt production decrease (**Fig. 8d**). Overall, magmatism governs thermal weakening and strain localization, while deformation migration regulates melt production. This interplay induces cyclic weakening and hardening of fault zones, shaping multiple detachment systems across the northern SCS margin (**Fig. 1b**).

*4.3 P-T-t path and tectonic subsidence*

We track five particle tracers exposed along the detachment fault to assess *P-T-t* paths (**Fig. 9a**, **Supplementary Movie S4**). During the early rifting stage (0–12.6 Myr), all tracers undergo near-isothermal decompression, with the shallow blue and orange tracers cooling along a geothermal gradient of 65°C/km from ~11 Myr. Following melt intrusion (from 12.6 Myr onward), the deep red and purple tracers exhibit retrograde heating (i.e., reheating during exhumation), whereas the middle green tracer remains unaffected and cools along a 100°C/km geothermal gradient. The red tracer temperature increases from 480°C at 13 Myr to a peak of 530°C at 15.8 Myr. Similarly, the purple tracer, initially at the upper/lower crust interface (560°C), reaches 740°C at 18 Myr, coinciding with the climax of the magma-detachment cycle (**Figs. 7b, 8**). Such heating process resembles the conceptual core complex model proposed by Lister and Baldwin (1993), which involves transient high-temperature pulses associated with magmatism. After the heating phase, the red and purple tracers are exhumed in response to tectonic denudation and undergo rapid cooling along a 100°C/km gradient. By the end of the local detachment cycle (~20 Myr), all tracers are carried upward to near-surface levels and cool to ≤ 250°C, after which surface sedimentation dominates the *P-T* conditions, leading to burial and increased pressure. To date, direct petrological data revealing *P-T* paths in the deep-water SCS margin are lacking. *P-T-t* paths from well-studied core complexes in the North American Cordillera and Aegean Sea manifest diverse exhumation styles (**Fig. 9a**). Nevertheless, our results share key first-order features with these observations, i.e., isothermal decompression followed by rapid cooling. While the numerical model of Huet et al. (2011) reproduced the *P-T* paths of several core complexes in the Cyclades, it failed to capture the heating phase recorded in Naxos (Buick and Holland, 1989). Our models demonstrate that this discrepancy could be reconciled by heating due to magmatic emplacement that accompanied post-orogenic extension in Naxos, a process analogous to that in the SCS (**Fig. 9a**). However, this interpretation remains tentative as some core complexes lack geochronological evidence for substantial syn-tectonic magmatism (Whitney et al., 2013). Alternative mechanisms include shear heating, advective heat transport by fluids, and enrichment of radiogenic elements (Lamont et al., 2023, and references therein).

We calculate the tectonic subsidence for five tracers to assess the effect of magma intrusion (**Fig. 9b**). During early stretching, the sub-basin subsides steadily at ~120 m/Myr. Following crustal magma intrusion at 12.6 Myr, subsidence in the basin depocenter (red and purple tracers) accelerates rapidly, peaking at 1,000 m/Myr. This increase comes from intense strain localization from magmatic thermal weakening, which enhances crustal thinning (Buck, 2004; Shi et al., 2005). Continued extension leads to isostatic rebound, uplifting the lower crust and exposing a domed footwall. Subsidence is then punctuated by a short-period flexural uplift. As deformation and magma emplacement migrate, the abandoned detachment fault ceases to accommodate strain, leading to a marked reduction in subsidence rate after ~19.5 Myr (**Fig. 9b**). This pattern of rapid subsidence followed by declining subsidence rates is consistent with back-stripping results from wells and seismic data in the Baiyun and Kaiping sags of the northern SCS margin (Shi et al., 2005; Xu et al., 2024). Note that in a static situation, magma emplacement can cause both subsidence and uplift. Uplift occurs when magma underplates beneath the Moho, replacing denser mantle with magma, while subsidence results from magma intrusion into the crust, where less dense rocks are replaced by denser intrusions. In the magma-rich margins of the North Atlantic, significant magmatic underplating is estimated to produce uplift of around 300–400 m, followed by minor subsidence as the magma cools, resulting in overall uplift (Maclennan and Lovell, 2002). Intermediate margins differ, with less voluminous underplating and more prominent faulting. In our dynamic model, magma intrusion into the crust leads to subsidence and magmatic thermal weakening leads to further localization of strain on the overlying detachments, causing, in turn, further thinning and up to 3 km of subsidence.

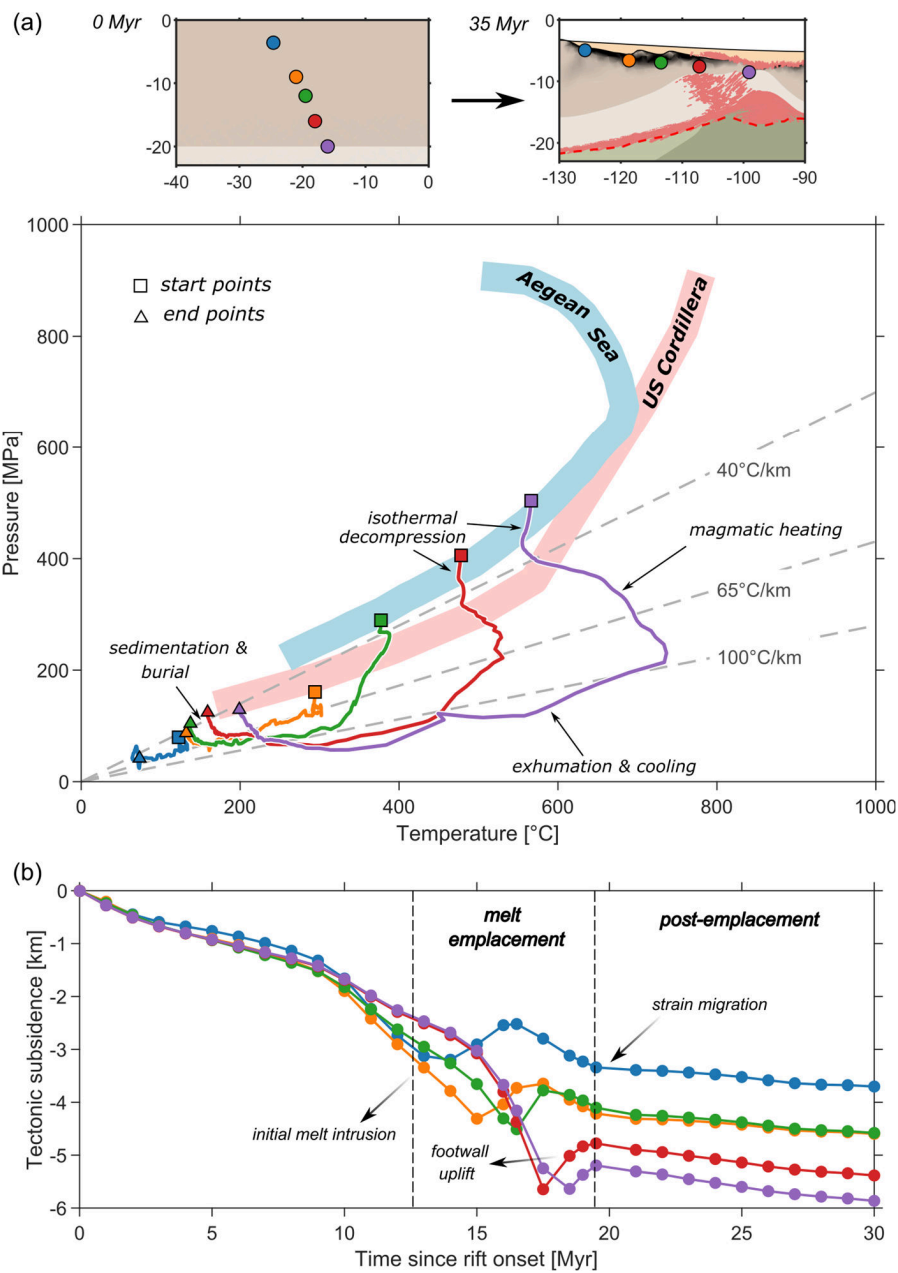

Fig. 9. (a) Pressure-temperature (*P-T*) paths of five tracers along detachment fault in the reference model. Path colors correspond to positioned tracers (circles) in the top schematic. Observed natural datasets from Naxos in the Aegean Sea (Buick and Holland, 1989) and the Ruby-East Humboldt Range in the North American Cordillera (McGrew et al., 2000) are shown as colored bands. All tracers record near-isothermal decompression during early rifting, retrograde heating during melt intrusion (only deeper tracers), and rapid cooling during exhumation (also see Supplementary Movie S4). The results reproduce first-order features of core complexes worldwide and highlight the role of magmatic emplacement in heating process. (b) Tectonic subsidence history of five tracers in (a). Subsidence in the basin depocenter is initially enhanced by melt emplacement but subsequently hindered by the uplift of core complex system, reflecting dynamic feedback between magmatism and detachment activity.

### *4.4 Deformation mode at intermediate passive margins*

Rifted margins are increasingly recognized to span a spectrum between magma-poor and magma-rich end-members, with many showing intermediate characteristics shaped by complex feedbacks among inheritance, sedimentation, magmatism and hydrothermal processes (Pérez-Gussinyé et al., 2023). In particular, the relative timing of crustal thinning, mantle melting, and serpentinization plays a key role in determining margin evolution (Tugend et al., 2020). Here we compare rifted margin evolution at intermediate margins and magma-poor margins, exemplified by the SCS and the WIM, in order to highlight the key processes and factors that result in their differing crustal and magmatic architecture (**Fig. 10**).

The initial lithospheric thermal structure plays a crucial role in rifted margin style (**Figs. 5a, 5b & 5c**) (Brune et al., 2017; Peron-Pinvidic et al., 2019; Pérez-Gussinyé et al., 2023). In the southern North Atlantic, ancient lithospheric inheritance, dating back to the Neoproterozoic, is suggested to play a persisting role in subsequent tectonism, spanning multiple Wilson cycles (Welford, 2025). During the Paleozoic, the Variscan orogeny resulted from the closure of a series of narrow, immature oceans, involving primarily mechanical processes with limited thermal modification and fertile underlying mantle, which has a high potential for magma production (Chenin et al., 2019). However, the subsequent continental rifting is devoid of large magmatic activity, resulting in a notably magma-poor margin characterized by brittle sequential faulting and eventual mantle exhumation at the COT (Ranero and Pérez-Gussinyé, 2010). This extension and breakup style could be attributed to the inherited cold and strengthened orogenic lithosphere (Pérez-Gussinyé et al., 2001), resulting from the long post-orogenic thermal relaxation (> 60 Myr) from the end of Variscan orogeny to Mesozoic Atlantic rifting. In contrast, the SCS evolved from a magmatic arc, formed by northward subduction of the Paleo-Pacific in the Late Mesozoic (180– 66 Ma), into a rifted continental margin during Cenozoic rifting (45 Ma), driven by the southward subduction of the Proto-SCS (Hall, 2002). The short thermal relaxation and associated high geotherm, inherited from subduction-related magmatism and fluid activity, significantly influenced subsequent continental rifting, resulting in a wide rifted margin marked by a sharp COT zone without mantle exhumation (Larsen et al., 2018; Zhang et al., 2024). Particularly, our simulations, incorporating high lithospheric temperatures but ultra-slow extension rates similar to that in WIM, predict normal oceanic crust formation following continental breakup. In contrast, models of magma-poor margins with the same extension rates, but a colder initial thermal structure and thicker lithosphere, result in mantle exhumation at the continent-ocean transition (COT) (García-Pintado and Pérez-Gussinyé, 2025). In SCS models, continued extension at ultra-slow rates increases the brittle-ductile transition depth in distal margins (see Supplementary **Movie S4**). However, the combination of an inherited high geotherm and slow extension promotes magmatic underplating and intrusion already during rifting, which further heats and weakens the ductile lower crust. This process inhibits complete crustal embrittlement, serpentinization, and mantle exposure. Continuing rifting and magmatism result in enough magma production at break-up to form a normal-thickness magmatic oceanic crust directly adjacent to the distal margin (e.g., Model 4, **Fig. 7**), enabling a rapid transition to seafloor spreading as observed in seismic and drilling data (Larsen et al., 2018). Alternatively, Zhang et al. (2024) suggest that the rifting/spreading rate at breakup may further modulate margin style, noting that the initial seafloor spreading rate in the WIM is ultra-slow (< 20 mm/yr), whereas it accelerates up to ~40 mm/yr in the SCS. They propose a two-axis classification scheme based on magmatic budget and strain rate, rather than a simple binary end-member. Such variable extension rate, however, is held constant in our models and is not explicitly tested. Our models show that ultra-slow extension can lead to both a rapid transition to oceanic spreading or mantle exhumation, depending on inherited lithospheric structure.

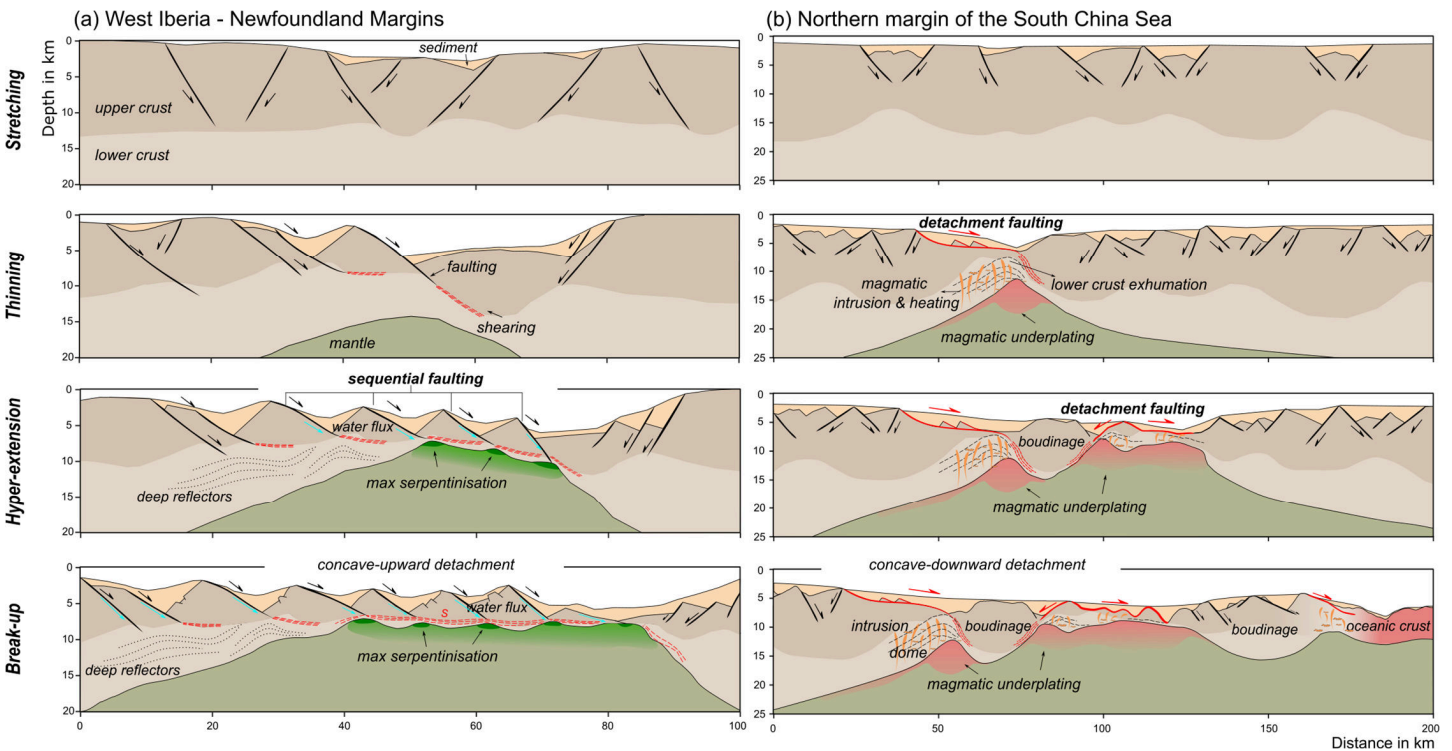

Fig. 10. Comparison of the crustal extension and detachment deformation in magma-poor margins and intermediate margins. (a) West Iberia–Newfoundland Margins, modified from Liu et al. (2022). An initially cool lithospheric structure leads to a predominance of brittle deformation. Large faults dip seaward and are sequentially active, generating large subsidence. Heating of the hanging wall enables ductile deformation at shallow crustal level, linking high-angle brittle faults and low-angle shear zones which localize strain from the upper crust to mantle. As new oceanward high-angle faults form sequentially, footwall rotation progressively flattens the shear zone,

leading to a detachment-like structure in the distal margin. Mantle serpentinization occurs through brittle deformation as the footwall cools. (b) Northern SCS margin. An initial warm and weak lithosphere, inherited from the supra-subduction setting, leads to both continent- and ocean-ward dipping faults and lower crustal extension by pure shear distributed in a wide area. Increasing extension results in melting and the emplacement of magma at the footwalls of active faults. Magmatic heat release promotes strain localization, causing initially steep, high-angle faults to rotate to lower angle through isostatic rebound of the footwall, forming single, long-lived detachment faults with concave-downward geometry. Oceanward rift migration is driven by asymmetric mantle uplift towards the hanging wall of large detachments, where the newer deformation will occur (Brune et al., 2017). Exhumation of the ductile lower crust leaves the competent upper crust to develop boudinage structures flanked by detachment systems. Strain re-localization and increased melting lead to break-up, followed by the formation of a magmatic oceanic crust. The detachment and associated core complex systems are observed in the proximal, distal and COT sections of the northern SCS margin. Note that the horizontal scale in (b) is twice that of (a).

Although large low-angle detachment structures are observed in both the SCS and WIM, their formation mechanisms and structures differ significantly. In the WIM, the 'S reflectors' from seismic reflection profiles were described as a listric detachment-like geometry at the interface between brittle crustal blocks and serpentinized mantle (Pérez-Gussinyé and Reston, 2001; Lymer et al., 2019). The formation of this structure was previously suggested to be aided by the serpentinization-induced weakening in the brittle field (Pérez-Gussinyé and Reston, 2001; Lymer et al., 2019). However, recent kinematic-dynamic modeling demonstrates that it may be formed by the cumulative action of various sequentially activated faults whose deepest segments rotate to low angles in the ductile field and amalgamate into a single reflector (Liu et al., 2022) (**Fig. 10a**). Simultaneously, the cooling of footwall under low extension velocity allows fluid flux along the active faults, resulting in mantle serpentinization with very little magmatic products (Liu et al., 2022). In contrast, the large detachment systems in the northern SCS margin are accompanied by distributed upper crustal boudinage and lower crust exhumation structures (Deng et al., 2020; Xu et al., 2024), which our models attributed to syn-tectonic magmatism driven by mantle partial melting (**Fig. 10b**). The geometry and kinematics of detachment faults resemble the 'rolling-hinge' model (Buck, 1988), wherein a single steeply dipping normal fault rotates to shallower dips through isostatic rebound of a hot, ductile footwall (**Fig. 8**). However, in the WIM scenario, oceanward sequential faulting does not nucleate in the same region, preventing successive slivers of crust from being consistently sliced from a fixed hinge (Ranero and Pérez-Gussinyé, 2010). Additionally, the detachment geometry in the distal domain of the WIM continental margin is concave-upward (listric), whereas in the SCS it is concave-downward, which is less effective in accommodating subsidence (**Fig. 9b**), and is more akin to core complexes in oceanic settings. These differences can be explained by the thermal weakening and prolonged strain localization facilitated by footwall magmatic intrusions, which occur as a consequence of rifting in a hotter environment in the SCS. In this scenario, thermal and mechanical feedback from magmatism acts to pin the location of a single, long-lived detachment, preventing the abrupt cooling and strengthening that would otherwise force fault abandonment and rift migration.

These conditions allow for sustained footwall exhumation, in contrast to the transient, sequential faulting of magma-poor systems.

## 5. Conclusion

We use 2D numerical modeling to investigate the interplay between tectonic extension and magmatism during continental rifting at intermediate margins, exemplified here by the SCS. Our simulations reproduce key features of wide rifted margins, including hyper-extended crust, boudinage structures, large-scale detachment fault systems and associated domed core complexes, which are characteristic of the northern continental margin of the SCS. We show that the formation of detachment faults and core complexes is fundamentally controlled by ductile flow in the lower crust, which is promoted by coeval footwall magmatic intrusions resulting from mantle decompression. Magmatic heat release promotes thermal weakening, enabling prolonged strain localization and the development of stable, long-lived detachment systems. In contrast, the polyphase, out-of-sequence faulting is dominant in the scenarios lacking magmatic heating. Melt emplacement initially accelerates tectonic subsidence by enhancing crustal thinning but this is later suppressed by doming associated with detachment structures.

We conclude that the inherited lithospheric thermal and rheological structure and the magnitude of syn-rift magmatism critically shape rift architecture. In the WIM, which evolved from a cold, thermally stable lithosphere, deformation is dominated by the lack of syn-rift magmatism, brittle sequential faulting, concave-upward detachment formed by rotation and amalgamation of deep ductile fault segments, mantle serpentinization supported by crustal embrittlement, and final mantle exhumation at the COT. In contrast, the northern SCS margin represents an intermediate rifting style, inherited from a warm and weak, subduction-modified lithosphere. Here, magmatic heating facilitates the formation of concave-downward detachment faults, exhumation of ductile lower crust, and boudinage structures. Faulting localizes around a fixed rolling hinge, with delayed abandonment due to persistent thermal weakening. The elevated initial lithospheric temperature and mantle partial melting result in the formation of normal oceanic crust after breakup, even in conditions of ultra-slow extension similar to those at magma-poor margins. These findings highlight a key distinction between magma-poor and intermediate margins, emphasizing the critical role of magmatism in controlling fault kinematics, crustal architecture, and the eventual breakup style.

Supplementary Materials for

# Interactions between syn-rift magmatism and tectonic extension at intermediate rifted margins

Peng Yang, Marta Pérez-Gussinyé, Shaowen Liu, Javier García-Pintado and Gudipati RaghuRam

**Contents of this file**

**Additional Supporting Information (Files uploaded separately)**
Movies S1 to S4

*Text S1. Governing equations*
Rift2Ridge is a 2D numerical code to simulate continental rifting and seafloor spreading. This code originally branched from MILAMIN (Dabrowski et al., 2008). It has previously been used to investigate a broad spectrum of rifting processes, e.g., the implications of continental rifting style on melting and serpentinization (Ros et al., 2017), the influence of surface processes — i.e., erosion and sedimentation — on the architecture and evolution of passive margins during continental rifting (Andrés-Martínez et al., 2019), variations in sedimentation and unconformity patterns at continental margins during rifting and breakup (Pérez-Gussinyé et al., 2020), estimation of hydrogen production during mantle exhumation at the Continental-Oceanic Transition zones of continental passive margins (Liu et al., 2023; García-Pintado and Pérez-Gussinyé, 2025), the interplay between cratonic roots and rifted margin asymmetry (Raghuram et al., 2023), and the relationship between tectonics, mantle flow and melting at ultra-slow oceanic spreading ridges (Mezri et al., 2024). For completion, the main code components are summarized here. The code uses the finite element method to solve the equations of momentum, mass and energy conservation. Deformation and pressures are calculated in a Lagrangian grid by solving the Stokes force-balance equation assuming incompressible condition:

$$\nabla \cdot \tau - \nabla P + \rho g = 0 \tag{1}$$

where $\tau$ is the deviatoric stress, $P$ the total pressure, $\rho$ the density, and $g$ the gravity acceleration. The mass conservation equation is:

$$\nabla \cdot v = 0 \tag{2}$$

where v is the velocity vector. Temperature is calculated by solving the thermal energy conservation equation:

$$\rho C_p \cdot \frac{\partial T}{\partial t} = \nabla \cdot (k \nabla T) + Q \tag{3}$$

where $C_p$ is the effective heat capacity, $T$ is the temperature, $t$ is the time, $k$ is the thermal conductivity, and $Q$ is the heat sources. In our model the heat sources include five components:

$$Q = Q_r + Q_p + Q_s + Q_{ml} + Q_{ms} \tag{4}$$

where $Q_r$ is the crustal radioactive heat production, $Q_p$ is shear heating production which depends on stress and strain rate, $Q_s$ is heat generation by serpentinization reaction, and $Q_{ml}$ and $Q_{ms}$ are the latent and sensible heat release from melt emplacement. After jointly solving for the Stokes and mass balance conservation equations, the solution for the heat conduction equation follows. The thermal energy is advected with the Lagrangian moving mesh.

*Text S2. Rock rheology*

Rock mechanical behavior is modelled with a visco-elasto-plastic rheology with an additive decomposition of the deviatoric strain rate:

$$\tau = \eta_{eff} \left( 2\dot{\varepsilon}' + \frac{\tau^{oldJ}}{\mu \Delta t} \right) \tag{5}$$

where $\tau$ is shear stress, $\eta_{eff}$ is the effective viscosity, $\dot{\varepsilon}'$ is the deviatoric strain rate, $\tau^{oldJ}$ is the Jaumann corotational stress, $\mu$ is the shear modulus and $\Delta t$ is the numerical time step. The effective viscosity formulation depends on the deformation regime, namely whether the material undergoes plastic yielding or deforms in the visco-elastic regime. We implement a Drucker-Prager yield criterion to recognize where plastic deformation occurs, that is material is considered to deform plastically when the square root of the second invariant of the deviatoric stress, $\tau_{II}$, is larger than or equal to the yield stress ($\tau_{II} \geq \sigma_{yield}$) (Ranalli, 1995; Moresi et al., 2003), where $\sigma_{yield}$ is:

$$\sigma_{yield} = P sin\varphi + C cos\varphi \tag{6}$$

where $C$ is the cohesion and $\varphi$ is the friction angle. Using Prandtl-Reuss flow law, the plasticity can be included into viscous formulation (e.g., Andrés-Martínez et al., 2019) and the effective viscosity $\eta_{eff}$ is defined as:

$$\eta_{eff} = \frac{\sigma_{yield}}{\left(2\dot{\varepsilon}' + \frac{\tau^{oldJ}}{\mu\Delta t}\right)_{II}} \qquad (7)$$

Otherwise, visco-plastic deformation dominates the model ($\tau_{II} \leq \sigma_{yield}$), so that the effective viscosity is:

$$\eta_{eff} = \left(\frac{1}{\eta_{dis}} + \frac{1}{\eta_{dif}} + \frac{1}{\mu\Delta t}\right)^{-1} \qquad (8)$$

where $\eta_{dis}$ and $\eta_{dif}$ are the dislocation and diffusion creep viscosity, respectively. The viscous flow is described by nonlinear power-law creep rheology, such that:

$$\eta_{dis/dif} = SB^{-\frac{1}{n}}\dot{\varepsilon}_{II}^{\frac{1-n}{n}} exp\left(\frac{E + PV}{nRT}\right) \qquad (9)$$

where $S$ is a factor for scaling parameters obtained from uniaxial/triaxial experiments to the second-invariant-based formulation of the viscosity, $B$ is the pre-exponential factor of the flow law, $n$ is the power-law exponent, $E$ is the activation energy, $V$ is the activation volume, $R$ is the gas constant and $T$ is the absolute temperature.

*Text S3. Softening mechanism and weak seed implementation*

Faults and shear zones acquire a lower effective viscosity and tend to be weakened as they undergo deformation. Weakening mechanisms in the brittle regime include loss in rock cohesion along faults and fractures with accumulated strain, and effective friction angle reduction resulting from fluid pressure increase and mineral transformation with weak mineral replacement (Bos and Spiers, 2002; Lavier et al., 2000). Here we implement strain softening by reducing both friction angle and rock cohesion. The friction angle can be described by a linear parametric function:

$$\theta = \theta_0 + \frac{\theta_1 - \theta_0}{E_1 - E_0}(E - E_0) \qquad (10)$$

where $E$ is the second invariant of the accumulated plastic strain ($E_0 = 0, E_1 = 1$), and $\theta$ is the calculated friction angle used for solving plasticity. We set the initial friction angle $\theta_0$ as 30° (friction coefficient of 0.577) and reduce it to $\theta_1$ as 15° (friction coefficient of 0.268) when the accumulated plastic strain increases from 0 to 1. Cohesion reduction is parametrized in a similar way, where we set $C_0$ = 10 MPa, and $C_1$ = 4 MPa. The equivalent equation is also used in the viscous strain softening where the pre-exponential factor $B$ is multiplied by 1 when the accumulated viscous strain is 0 and by 15 when the strain reaches 1. This implementation is limited to temperatures > 800 °C and strain softening progressively vanishes at temperatures of 1200 °C (see Ros et al., 2017; Raghuram et al., 2023).

We use a random damage weak seed at the beginning to localize the initial deformation (see Pérez-Gussinyé et al., 2020). This is achieved by changing the initial friction angle $\theta_0$ randomly with a maximum amplitude of 4°, so the initial friction angle varies between 28° and 32°. This perturbation follows a horizontal-space Gaussian function (100 km standard deviation) decaying from the center to the sides of the model domain.

*Text S4. Surface processes*

To simulate the sediment transport, we use a 1-D diffusion equation based on conversation of mass to solve erosion and deposition (Andrés-Martínez et al., 2019; Pérez-Gussinyé et al., 2020):

$$\frac{\partial h}{\partial t} = \frac{\partial}{\partial x}\left((K + cq_w^n)\frac{\partial h}{\partial x}\right) \quad (11)$$

where $h$ is the topography, $x$ is horizontal distance, $K$ is the slope diffusivity of the sediments, $c$ is the transport coefficient, $q_w$ is the water flux, and $n$ is the power law which represents the type of relationship between the sediment transport and the water flux ($n \geq 1$). The water flux $q_w$ at downstream locations can be defined as:

$$q_w = \alpha x_d \quad (12)$$

where $\alpha$ is the water discharge/effective rainfall, and $x_d$ is the integrated downstream distance from topographic highs to the valley bottom.

In the submarine domain, sediment transport is affected by the motion of wave and tide, which weaken with increasing water depth. The sediment transport can be described as:

$$\frac{\partial h}{\partial t} = \frac{\partial}{\partial x}\left(K_s e^{(-\lambda_s h_w)}\frac{\partial h}{\partial x}\right) + S \quad (13)$$

where $K_s$ is the submarine diffusion coefficient, $\lambda_s$ is the submarine diffusion decay coefficient, $h_w$ is the water depth, and $S$ is the hemipelagic sediment rate. The parameters of the erosion and sedimentation processes are shown in **Table S1**. The effect of sedimentation and erosion on margin architecture and the effect of margin evolution on sedimentary stratigraphy were described comprehensively in previous studies (Andrés-Martínez et al., 2019; Pérez-Gussinyé et al., 2020), which are not included in this work.

*Text S5. Melting processes*

*S5.1 Partial melting of mantle*

Partial melting is calculated following Phipps Morgan (2001), where melting is assumed to be a reversible adiabatic process. Partial melting starts when the new temperature exceeds the dry pressure-dependent and depletion-dependent solidus temperature:

$$T_s = T_{s_0} + \frac{\partial T_s}{\partial P}P + \frac{\partial T_s}{\partial F}F \quad (14)$$

where $T_{s_0}$ = 1081 °C is the solidus temperature at the surface, $\frac{\partial T_s}{\partial P}$ = 132 °C/Pa is the solidus-pressure dependence, $\frac{\partial T_s}{\partial F}$ = 350 °C is the solidus-depletion dependence. $F \in [0, 1]$ is melt depletion, i.e. the ratio between extracted melt and solid residue. In our formulation, the mantle temperature in the melting area reduces to the solidus temperature as latent heat of melting is consumed. The melt productivity $\Delta F$ is given by:

$$\Delta F = \frac{\Delta T}{\frac{L}{C_p} + \frac{\partial T_s}{\partial F}} \tag{15}$$

where $L$ is the latent heat of melting, $\frac{L}{C_p}$ = 550 °C, and $\Delta T$ is the difference between the temperature of rock and solidus. At each time step, the produced melt yields a total extracted melt volume $V_m$ (m³) across the model domain:

$$V_m = 0.8 \times \int_\Omega \Delta F_m \tag{16}$$

where $\Delta F_m$ is the produced melt fraction in this time step. The multiplier 0.8 represents that only 80% of the total melt volume is extracted and the remaining part refertilizes the mantle (Behn and Grove, 2015)

The density and viscosity of the mantle are updated considering the effect of depletion. The density is calculated following the Boussinesq approximation:

$$\rho = \rho_0(1 - \alpha(T - T_0) - \beta F) \tag{17}$$

where $\rho$ is the density, $\alpha$ is the thermal expansivity, $\beta$ is a factor that parameterizes the effect of depletion on density ($\beta$ = 0.044 (Armitage et al., 2013)), $F$ is the melt depletion, $\rho_0$ is the reference density (see **Table S1**), and $T_0$ is the reference temperature (0 °C).

The effect of depletion on model composition and viscosity is considered by using a mixed rheology. The initial asthenospheric mantle is dry olivine with 500 ppm H/Si without depletion (Hirth and Kohlstedt, 2003). As mantle melting occurs and the depletion increases from 0% to 4%, the rheology and viscosity transfers linearly from wet olivine to dry olivine. Beyond 4%, all water is assumed to be extracted (Phipps Morgan, 1997), and the rheology becomes entirely dry olivine.

*S5.2 Heat release from melt emplacement*

The emplacement of generated melt is represented by tracking points, which after emplacement are advected with the velocity fields along subsequent model steps. Sills are modeled as six elliptical bodies carrying equal volume arranged in two rows and three columns, with each ellipse comprising multiple tracking points. Similarly, dykes are rectangular, with height computed as $h = V/w$, where $V$ is melt volume and $w$ is dyke width, given by $w = v \times dt$ ($v$ is the full-extension velocity and $dt$ is the time step). We assign a reference density of 3000 kg m⁻³ to solidified magma, based on empirical density-$V_p$ relationships and consistent with high-velocity lower crustal bodies with $V_p$ of 7.0–7.8 km s⁻¹ (Brocher, 2005).

Heat release of melt emplacement is estimated following Mezri et al. (2024). As magma is emplaced at shallow levels heat released will come from the latent heat of crystallization and the sensible heat. Sensible heat results from the temperature difference between the host rock and emplaced magma. The total energy increment from sensible heat release $\Delta E_{ms}$ is calculated as:

$$\Delta E_{ms} = \int_{\Omega_m} \Delta\varepsilon_{ms} \approx \int_{\Omega_m} \rho_m C_m (T_m - T^b) \approx \sum_i^{n_m} v_i \rho_{m_i} C_{m_i} (T_{m_i} - T_i^b) \tag{18}$$

where $\Omega_m$ is the volume of emplaced melt, $\Delta\varepsilon_{ms}$ is the energy density (J m$^{-3}$), $\rho_m$ is the density of emplaced melt, $C_m$ is the heat capacity of emplaced melt, $T_m$ is melt emplacement temperature, $T^b$ is the background temperature field of host rocks, $v_i$ is the volume of tracking points used to represent dykes and sills, $n_m$ is the number of the tracking points, $i$ is the tracking point index. Here we set $T_m$ = 1100 °C (Sen, 2013), $\rho_m$ = 2800 kg m$^{-3}$, $C_m$ = 1200 J kg$^{-1}$ K$^{-1}$. As shown in **Fig 5**, discontinuities in the evolution of sensible heat mark the termination of an old detachment and the onset of a new one.

As for the latent heat release, the total energy $\Delta E_{ml}$ is calculated as:

$$\Delta E_{ml} = \int_{\Omega_m} \Delta\varepsilon_{ml} \approx \int_{\Omega_m} \rho_m \Delta H_{ml} \approx \sum_i^{n_m} v_i \rho_{m_i} \Delta H_{ml} \tag{19}$$

where $\Delta H_{ml} = 4 \times 10^5$ J kg$^{-1}$ is the enthalpy of crystallization (Sleep and Warren, 2014).

To consider the temperature-dependent kinetics of melt crystallization, we use a simple exponential model that governs the progressive release of latent heat during solidification. The crystallization timescale is defined as:

$$\tau_{\text{cryst}} = \frac{1000}{log(10^7 \cdot t)} \tag{20}$$

where $t$ is one year, expressed in seconds. This equation assumes that for a host rock with temperature of 1000 °C, the melt requires 10 Myr to fully crystallize. The crystallization time at a given temperature $T$ is then parameterized as:

$$t_{\text{cryst}}(T) = \exp\left(\frac{T}{\tau_{\text{cryst}}}\right) \tag{21}$$

During each model timestep $\Delta t$, only a fraction of the available latent heat is released, proportional to the degree of crystallization that occur within that time frame. The fraction is computed as:

$$f_{\text{cryst}} = min\left(1, \frac{\Delta t}{t_{\text{cryst}}(T)}\right) \tag{22}$$

The energy associated with latent heat release is then redefined as $\Delta E_{ml\_new} = \Delta E_{ml} \cdot f_{\text{cryst}}$.

As mentioned in the main paper, here we choose to place 1/3 of the total extracted melt beneath the top basement, while the remaining melt is portioned with 1/3 emplaced in the lower crust and 2/3 beneath the Moho. An alternative model is conducted to test the sensitivity of this assumption (Table S2, Model 7).

Finally, to map heat release into the finite element mesh, we distribute the energy sources from magma emplacement in a way that is inversely proportional to the distance between mesh nodes and points that track the emplaced melt. For this we determine a scalar factor $\widehat{\Delta\varepsilon_m}$ by solving

$$\widehat{\Delta\varepsilon_m} \int w = \Delta E_m \tag{23}$$

where $w$ is a dimensionless weight field that equals 1.0 within the perimeter of dykes or sills and linearly decays to 0.0 at a 500 m cut-off distance. $\Delta E_m$ represents either $\Delta E_{ms}$ or $\Delta E_{ml}$. The nodal heat source $\boldsymbol{Q}_{ms}$ and $\boldsymbol{Q}_{ml}$ are then computed for sensible and latent heat, respectively (see Mezri et al. (2024) for further information).

### Text S6. Hydrothermal cooling

We use a parameterized conductive hydrothermal cooling approach to account for the cooling effects of seawater circulation based on Morgan and Chen (1993), which assumes maximum hydrothermal cooling along active faults (see Mezri et al., 2024). This is achieved by increasing the thermal conductivity $k$ in faults and fracture networks where water infiltrates the basement:

$$k = Nu \cdot k_0 \qquad (24)$$

where $k_0$ is the initial base thermal conductivity, and $Nu$ is a dimensionless Nusselt number, which can be estimated by a deformation-dependent equation:

$$Nu = 1 + (Nu_{max} - 1)\left(1 - e^{\left(-\frac{3\dot{\varepsilon}_{II}^p}{\dot{\varepsilon}^H}\right)}\right) \qquad (25)$$

where $\dot{\varepsilon}_{II}^p$ is the second invariant of the plastic strain rate, used as a proxy for permeability, $\dot{\varepsilon}^H$ is a nominal plastic strain rate. Here we set $Nu_{max} = 8$, and $\dot{\varepsilon}^H = 1\times10^{-13}$ s$^{-1}$. This hydrothermal cooling is limited below 600 °C isotherm, the threshold where rock permeability becomes too low for fluid flow.

### Text S7. Model setup

The model lithosphere consists of three-layers: a wet quartzite upper crust (Gleason and Tullis, 1995), a wet anorthosite lower crust (Rybacki and Dresen, 2000), and a dry olivine lithospheric mantle (Hirth and Kohlstedt, 2003) (**Fig. 2a**). The asthenosphere is assigned a wet olivine rheology (Hirth and Kohlstedt, 2003), and a 5 km-thick transition zone at the lithosphere-asthenosphere boundary (LAB) contains a linear mixture of wet and dry olivine, with depletion decreasing from 4% to 0%.

Mesh resolution is set to 5 km at the model bottom, 4 km at the Moho, 2 km at the upper/lower crust interface, and 1 km at the surface. A local refinement to ~0.75 km is applied in a 200 km × 60 km upper zone. The time step is 10 kyr. Radiogenic heat production is set to 1.7 $\mu$W m$^{-3}$ for the upper crust, within the measured range of 1.65 to 2.0 $\mu$W m$^{-3}$ (Hu et al., 2020), and 1.3 $\mu$W m$^{-3}$ for the lower crust. A relatively high Moho temperature of 800 °C is then prescribed (**Fig. 2b**) to consider a non-steady state configuration representing a young thermal lithosphere (e.g., Burov and Diament, 1995). Such initial conditions are appropriate for the SCS, where the thermal structure is typical of back-arc regions and short post-subduction thermal relaxation (< 60 Ma).

The pre-rift crustal thickness is set to 35 km thick, with an upper-to-lower crust ratio of 20:15 km based on seismic data from southeastern China and less extended Pearl River Estuary Basin (Hayes and Nissen, 2005; Cao et al., 2014). The initial LAB depth is 100 km, consistent with heat flow and seismic estimates for onshore lithosphere (Nissen et al., 1995; An and Shi, 2006). A symmetrical half-extension velocity of 5 mm/yr is applied on both sides, following plate reconstruction of the SCS (Brune et al., 2016). The abrupt acceleration of extension before breakup, attributed to strength loss (Brune et al., 2016), is not considered. Thermal boundary conditions are imposed with a surface temperature of $T_0 = 0$ °C and a LAB temperature of $T_b = 1360$ °C, based on numerical studies of present-day mantle potential temperature beneath the SCS (Zhang et al., 2021). Detailed thermomechanical parameters are listed in Supplementary **Table S1.**

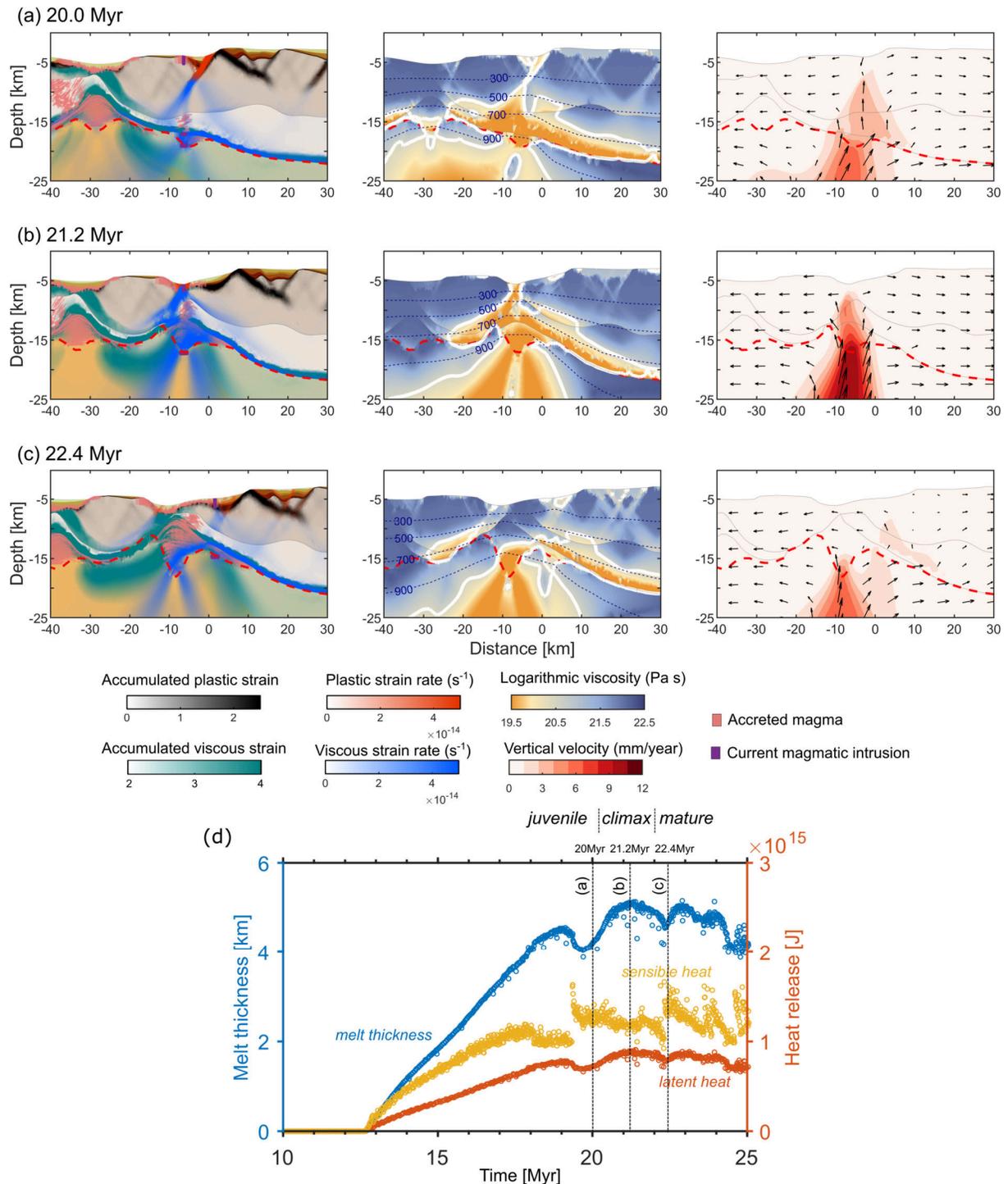

Fig. S2. Another detachment-magma cycle from ca. 20 Myr to 22.4 Myr of the reference model. Evolutionary cycle is divided into (a) juvenile, (b) climax and (c) mature stage according to the pattern of deformation and magmatism. (d) Melt thickness, sensible heat and latent heat through time. The times corresponding to the three evolutionary stages within the detachment life-cycle are shown with dashed lines. Here the detachment fault is dipping to the continent. See the main text (**Section 4.2**) for detailed explanation.

# Table S1. Thermomechanical parameters for numerical models

| Thermomechanical parameters | | | | |
|---|---|---|---|---|
| Variable [unit] | Wet quartzite (UC) | Wet anorthite (LC) | Dry olivine (LM) | Wet olivine (AM) |
| Dislocation pre-exponential factor $log\,(B_{dis})$ [Pa$^{-n}$ s$^{-1}$] | -28.0 | -15.4 | -15.96 | -15.81 |
| Dislocation exponent $n_{dis}$ | 4.0 | 3.0 | 3.5 | 3.5 |
| Dislocation activation energy $E_{dis}$ [kJ/mol] | 223 | 356 | 530 | 480 |
| Dislocation activation volume $V_{dis}$ [10$^{-6}$ m$^3$/mol] | - | - | 13 | 10 |
| Diffusion pre-exponential factor $log\,(B_{dif})$ [Pa$^{-n}$ s$^{-1}$] | - | - | -8.16 | -8.64 |
| Diffusion exponent $n_{dif}$ | - | - | 1 | 1 |
| Diffusion activation energy $E_{dif}$ [kJ/mol] | - | - | 375 | 335 |
| Diffusion activation volume $V_{dif}$ [10$^{-6}$ m$^3$/mol] | - | - | 6 | 4 |
| Shear modulus $\mu$ [GPa] | 36 | 40 | 74 | 74 |
| Thermal conductivity $k$ [Wm$^{-1}$ K$^{-1}$] | 2.4 | 2.5 | 3.4 | 3.4 |
| Hear capacity $C_p$ [J kg$^{-1}$ K$^{-1}$] | 1200 | 1200 | 1200 | 1200 |
| Radiogenic heat production $H$ [$\mu$W m$^{-3}$] (reference model) | 1.7 | 1.3 | 0 | 0 |
| Bulk density $\rho_0$ [kg m$^{-3}$] | 2700 | 2850 | 3360 | 3360 |
| Coefficient of thermal expansion $\alpha$ [10$^{-5}$ K$^{-1}$] | 2.4 | 2.4 | 3.0 | 3.0 |
| Cohesion (initial – final) $C_0$ [MPa] | 10 – 4 | | | |
| Friction angle (initial – final) $\phi_0$ [°] | 30 – 15 | | | |
| Surface process parameters | | | | |
| Subaerial hillslope diffusion $K$ [m$^2$/year] | 0.25 | | | |
| Subaerial discharge transport coefficient $\alpha$ | 10$^{-3}$ | | | |
| Submarine diffusion coefficient $K_s$ [m$^2$/year] | 10$^3$ | | | |
| Submarine diffusion coefficient decay $\lambda_s$ [m$^{-1}$] | 2.5 × 10$^{-2}$ | | | |
| Pelagic sediment rate S [m/year] | 3.5 × 10$^{-4}$ | | | |

Note: rheological parameters for the upper crust (UC), lower crust (LC), lithospheric mantle (LM), and asthenospheric mantle (AM) are from Wilks and Carter (1990), Gleason and Tullis (1995), and Hirth and Kohlstedf (2003). Other parameters are taken from Turcott and Schubert (2014).

**Table S2. Summary of parameters for different simulations**

Note: All the models share a full extension velocity of 10 mm/yr and a total crustal thickness of 35 km. Model 4 is our reference model, closely resembling the South China Sea continental margin.

| Model ID | Melt generation | Melt density | Melt heat release | Moho temperature (°C) | Upper crust thickness (km) | Ratio of melts in the upper crust |
|---|---|---|---|---|---|---|
| 1 | ✗ | ✗ | ✗ | 800 | 20 | 1/3 |
| 2 | √ | ✗ | ✗ | 800 | 20 | 1/3 |
| 3 | √ | √ | ✗ | 800 | 20 | 1/3 |
| **4** | √ | √ | √ | 800 | 20 | 1/3 |
| 5a | √ | √ | √ | **600** | 20 | 1/3 |
| 5b | √ | √ | √ | **740** | 20 | 1/3 |
| 6a | √ | √ | √ | 800 | **15** | 1/3 |
| 6b | √ | √ | √ | 800 | **25** | 1/3 |
| 7 | √ | √ | √ | 800 | 20 | **1/4** |

**Legends of Movies:**

**Movie S1:** Evolution of continental rifting and breakup in *Model 1*. The simulation was run with an initial crustal thickness of 35 km, a mantle temperature of 1360 °C, and a full extension rate of 10 mm/yr. Note that the mantle partial melting is not included. Model phases are color-coded as follows: light taupe: upper crust, off-white: lower crust, light green: lithospheric mantle, and orange: asthenospheric mantle. The blue lines show isotherms. Plastic and ductile strain rates are shown in red and blue transparency scales, respectively. Accumulated strain is shown in gray transparency scales. Surface sediments are colored by age using a rainbow colormap. Black arrows indicate the velocity field of lower crust and mantle. Details of model setup are provided in **Fig. 2**.

**Movie S2:** Evolution of continental rifting and breakup in *Model 2*. The simulation was run with an initial crustal thickness of 35 km, a mantle temperature of 1360 °C, and a full extension rate of 10 mm/yr. Note that the mantle partial melting is incorporated, but the effect of density and heat release of magma emplacement are not considered. Model phases are color-coded as follows: light taupe: upper crust, off-white: lower crust, light green: lithospheric mantle, and orange: asthenospheric mantle. The blue lines show isotherms. Plastic and ductile strain rates are shown in red and blue transparency scales, respectively. Accumulated strain is shown in gray transparency scales. Surface sediments are colored by age using a rainbow colormap. Black arrows indicate the velocity field of lower crust and mantle. Pink/red markers show emplaced magmatic intrusion and underplating. Details of model setup are provided in **Fig. 2**.

**Movie S3:** Evolution of continental rifting and breakup in *Model 3*. The simulation was run with an initial crustal thickness of 35 km, a mantle temperature of 1360 °C, and a full extension rate of 10 mm/yr. Note that the mantle partial melting and density effect of melt emplacement are incorporated, but the magmatic heat release is not considered. Model phases are color-coded as follows: light taupe: upper crust, off-white: lower crust, light green: lithospheric mantle, and orange: asthenospheric mantle. The blue lines show isotherms. Plastic and ductile strain rates are shown in red and blue transparency scales, respectively. Accumulated strain is shown in gray transparency scales. Surface sediments are colored by age using a rainbow colormap. Black arrows indicate the velocity field of lower crust and mantle. Pink/red markers show emplaced magmatic intrusion and underplating. Details of model setup are provided in **Fig. 2**.

**Movie S4:** [*Top panel*] Evolution of continental rifting and breakup in *Model 4*. The simulation was run with an initial crustal thickness of 35 km, a mantle temperature of 1360 °C, and a full extension rate of 10 mm/yr. Note that this model includes all the melting processes: mantle partial melting, as well as the thermal and density effects of magma emplacement. Model phases are color-coded as follows: light taupe: upper crust, off-white: lower crust, light green: lithospheric mantle, and orange: asthenospheric mantle. The blue lines show isotherms. Plastic and ductile strain rates are shown in red and blue transparency scales, respectively. Accumulated strain is represented by gray shading. Surface sediments are colored by age using a rainbow colormap. Black arrows indicate the velocity field within the lower crust and mantle. Pink markers show magmatic intrusion and underplating. Details of model setup are provided in **Fig. 2**. [*Lower left panel*] Temporal evolution of generated melt thickness, as well as sensible and latent heat released from magma emplacement. [*Lower right panel*] Pressure-temperature-time (P-T-t) paths of five tracers along the detachment fault in Model 4. Path colors correspond to the positioned circles shown in the top panel schematic.